\documentclass{aa}

\usepackage{subfig}

\usepackage{array} 
\usepackage{tabularx}
\usepackage{graphicx}
\usepackage[version=3]{mhchem}
\usepackage{txfonts}
\usepackage{natbib}  
\bibpunct{(}{)}{;}{a}{}{,} 

\begin{document}
\title{New chemical scheme for studying carbon-rich exoplanet atmospheres}
\authorrunning{Venot et al.}

\author{Olivia Venot\inst{1}, Eric H\'ebrard\inst{2}, Marcelino Ag\'undez\inst{3}, Leen Decin\inst{1}, Roda Bounaceur\inst{2}}

\institute{Instituut voor Sterrenkunde, Katholieke Universiteit Leuven, Celestijnenlaan 200D, 3001 Leuven, Belgium \and 
Laboratoire R\'{e}actions et G\'{e}nie des Proc\'{e}d\'{e}s, LRGP UMP 7274 CNRS, Universit\'{e} de Lorraine, 1 rue Grandville, BP 20401, F-54001 Nancy, France\\
Instituto de Ciencia de Materiales de Madrid, CSIC, C/Sor Juana In\'es de la Cruz 3, 28049 Cantoblanco, Spain \and
\email{olivia.venot@ster.kuleuven.be}}

\date{Received; accepted}


\abstract
{While the existence of more than 1800 exoplanets have been confirmed, there is evidence of a wide variety of elemental chemical composition, that is to say different metallicities and C/N/O/H ratios. Atmospheres with a high C/O ratio (above 1) are expected to contain a high quantity of hydrocarbons, including heavy molecules (with more than two carbon atoms). To correctly study these C--rich atmospheres, a chemical scheme adapted to this composition is necessary.}
{We have implemented a chemical scheme that can describe the kinetics of species with up to six carbon atoms (C$_0$-C$_6$ scheme\thanks{This chemical network is available for the community at the online database KIDA: Kinetic Database for Astrochemistry at http://kida.obs.u-bordeaux1.fr/models/.}). This chemical scheme has been developed with combustion specialists and validated by experiments that were conducted on a wide range of temperatures (300 -- 2500~K) and pressures (0.01 -- 100~bar).}
{To determine for which type of studies this enhanced chemical scheme is mandatory, we created a grid of 12 models to explore different thermal profiles and C/O ratios. For each of them, we compared the chemical composition determined with a C$_0$-C$_2$ chemical scheme (species with up to two carbon atoms) and with the C$_0$-C$_6$ scheme. We also computed synthetic spectra corresponding to these 12 models.}
{We found no difference in the results obtained with the two schemes when photolyses were excluded from the model, regardless of the temperature of the atmosphere. In contrast, differences can appear in the upper atmosphere (P$>\sim$1-10~mbar) when there is photochemistry. These differences are found for all the tested pressure-temperature profiles if the C/O ratio is above 1. When the C/O ratio of the atmosphere is solar, differences are only found at temperatures lower than 1000~K. The differences linked to the use of different chemical schemes have no strong influence on the synthetic spectra. However, with this study, we have confirmed C$_2$H$_2$ and HCN as possible tracers of warm C-rich atmospheres.}
{The use of this new chemical scheme (instead of the C$_0$-C$_2$) is mandatory for modelling atmospheres with a high C/O ratio and, in particular, for studying the photochemistry in detail. If the focus is on the synthetic spectra, a smaller scheme may be sufficient, because it will be faster in terms of computation time.}


\maketitle

\section{Introduction}

Observing spectra of transiting exoplanets has proven to be an excellent way of obtaining information on the chemical composition of their atmospheres. To date, only a few chemical species (H, H$_2$O, CO$_2$, CH$_4$, CO, HCN, K, Na) have been inferred thanks to this technique \citep[e.g.][]{charbonneau2002, desert2008, redfield2008, swain2009water, swain2009molecular, sing2011, madhu2014b}. Accordingly, one of the main goals of chemical models of exoplanet atmospheres is to find the elemental composition and the chemical processes that lead to the observed abundances, adjusting different parameters such as the metallicity, the C/O elemental abundance ratio, or the strength of vertical mixing \citep[i.e.][]{kopparapu2012photochemical, MillerRicciKempton2012, moses2013chemical, moses2013GJ436b, venot2014GJ3470b, agundez2014GJ436b}. 

The existence of carbon-rich exoplanets has been suggested for the first time by \cite{madhusudhan2011high} who analysed the Spitzer spectra of WASP-12b. Nevertheless, a C/O ratio $\ge$ 1 in the atmosphere of this planet is quite controversial. Several studies have found that the atmosphere of WASP-12b could either have a solar C/O ratio (0,54) or be C-rich \citep{crossfield2012, swain2013, mandell2013, stevenson2014, madhu2014b}. To clearly conclude on the composition of this planet, more observations with high precision are necessary, using for instance the Wide Field Camera 3 instrument on the Hubble Space Telescope \citep{WFC32012}.
However, \cite{Madhu2012} and \cite{moses2013chemical} studied the influence of the C/O ratio on the chemical composition of hot Jupiter atmospheres. They found that models with a C/O ratio $\sim$ 1 agreed with observational spectra of WASP-12b, XO-1b, and CoRoT-2b, which indicates that a wide variety of C/O ratios might be indeed possible in exoplanet atmospheres.

To study the chemical composition of carbon-rich atmospheres, photochemical models must use chemical schemes adapted to this carbon enrichment. When the C/O ratio increases, more complex carbon species are produced. Thus, carbon-rich atmospheres are expected to contain heavy hydrocarbons with significant abundances that need to be included in the chemical scheme used in the modelling. Most of chemical schemes (primarily constructed to study atmospheres with solar abundances) do not contain such complex species and can consequently give biased results. \cite{kopparapu2012photochemical} studied the atmosphere of WASP-12b with two different C/O ratios (solar and twice solar), but their chemical scheme only contained 31 molecules, the heaviest hydrocarbon being C$_2$H$_2$. Thus, the abundance of some hydrocarbons may be over- or underestimated because heavier molecules are absent from the chemical scheme. On the other hand, \cite{moses2013chemical} used a more complex chemical scheme (90 species), which extends up to C$_6$H$_6$, which allows describing the more important hydrocarbons.\\
Nevertheless, the question of the robustness and the completeness of chemical schemes is not addressed in many publications that examine photochemical models. The chemical schemes are constructed from reaction networks that were originally made for Jovian planets \citep{liang2003source, liang2004insignificance, line2010high, line2011thermochemical, moses2011disequilibrium, moses2013chemical} in which, in the best cases \citep[e.g. not for][]{liang2003source, liang2004insignificance}, high-temperature reactions have been added and forward reactions have been reversed to reproduce the thermochemical equilibrium in absence of out-of-equilibrium processes \citep{visscher2011quenching, venot2012chemical}.
\cite{zahnle2009atmospheric} modestly describe the method they used to construct their chemical scheme: reaction rates were taken from the NIST database\footnote{http://kinetics.nist.gov/kinetics}, and they chose by order of decreasing priority between reported reaction rates according to relevant temperature range, newest review, newest experiment, and newest theory. \cite{venot2012chemical} presented the first chemical scheme that was validated as a whole by experiment. Their chemical scheme was constructed in collaboration with combustion specialists and was experimentally validated on a wide range of temperatures (300 -- 2500 K) and pressures (0.001 -- 100 bar), making this one of the currently most reliable chemical schemes. It is able to describe the kinetics of species containing up to two carbon atoms.\\

The C$_0$-C$_2$ scheme may be insufficient, for studying atmospheres enriched in heavier hydrocarbons (with more than two carbon atoms). That is why we have developed an extended version of the chemical scheme of \cite{venot2012chemical}, with species containing up to six carbon atoms. In Sect. \ref{sec:model} we present this new chemical scheme and the grid of models we constructed to test it. In Sect. \ref{sec:Resultats} we present the results obtained with the C$_0$-C$_6$ scheme and compare them with those obtained with the C$_0$-C$_2$ scheme. In Sect. \ref{sec:Spectra}, we discuss the implications for interpreting transmission and emission spectra of exoplanet
atmospheres and identify species that could play the role of tracers for warm C--rich atmospheres.

\section{Model}\label{sec:model}
\begin{figure*}[t]
\centering
\includegraphics[angle=0,width=\columnwidth]{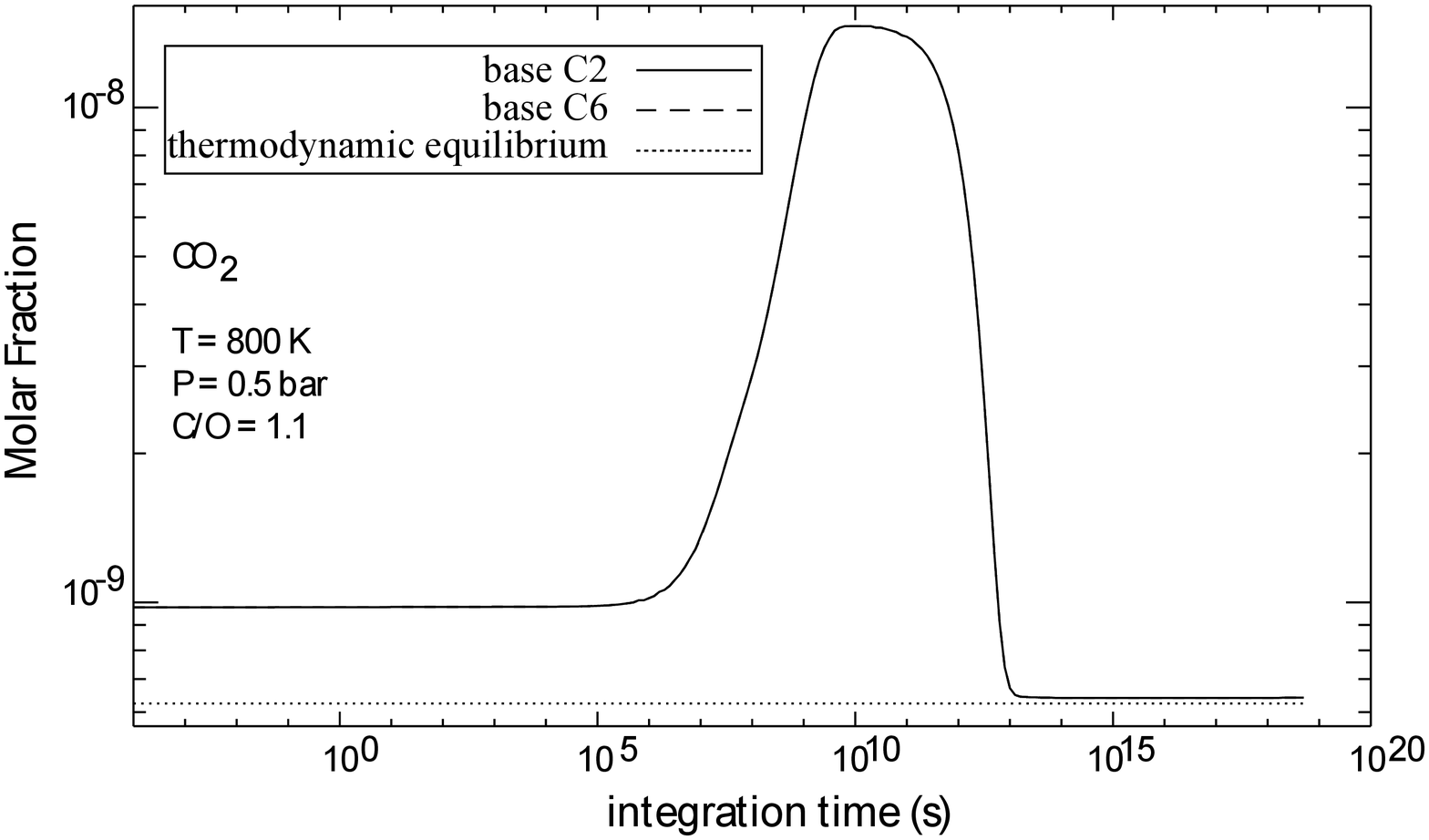} 
\includegraphics[angle=0,width=\columnwidth]{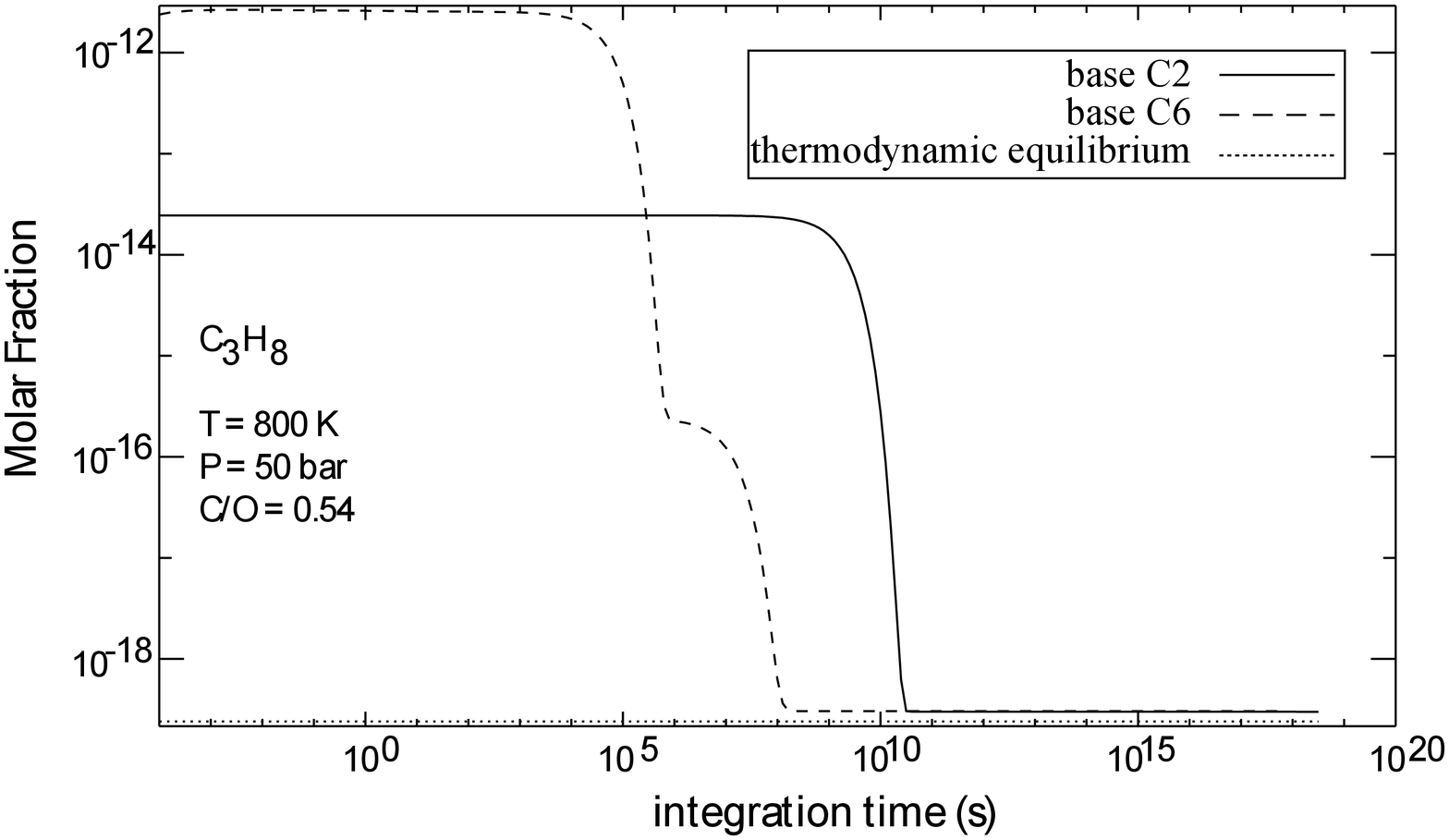}
\includegraphics[angle=0,width=\columnwidth]{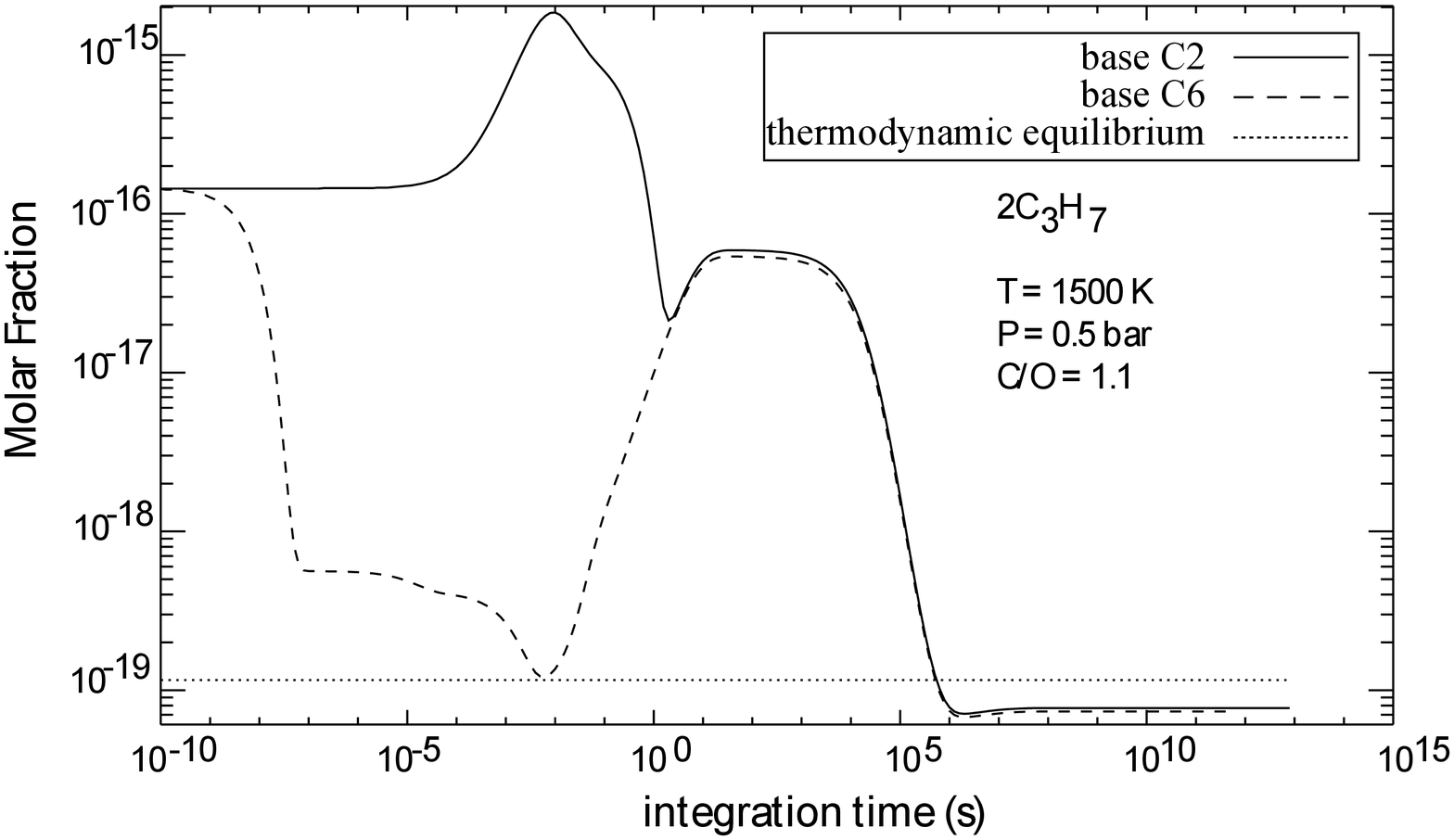}
\includegraphics[angle=0,width=\columnwidth]{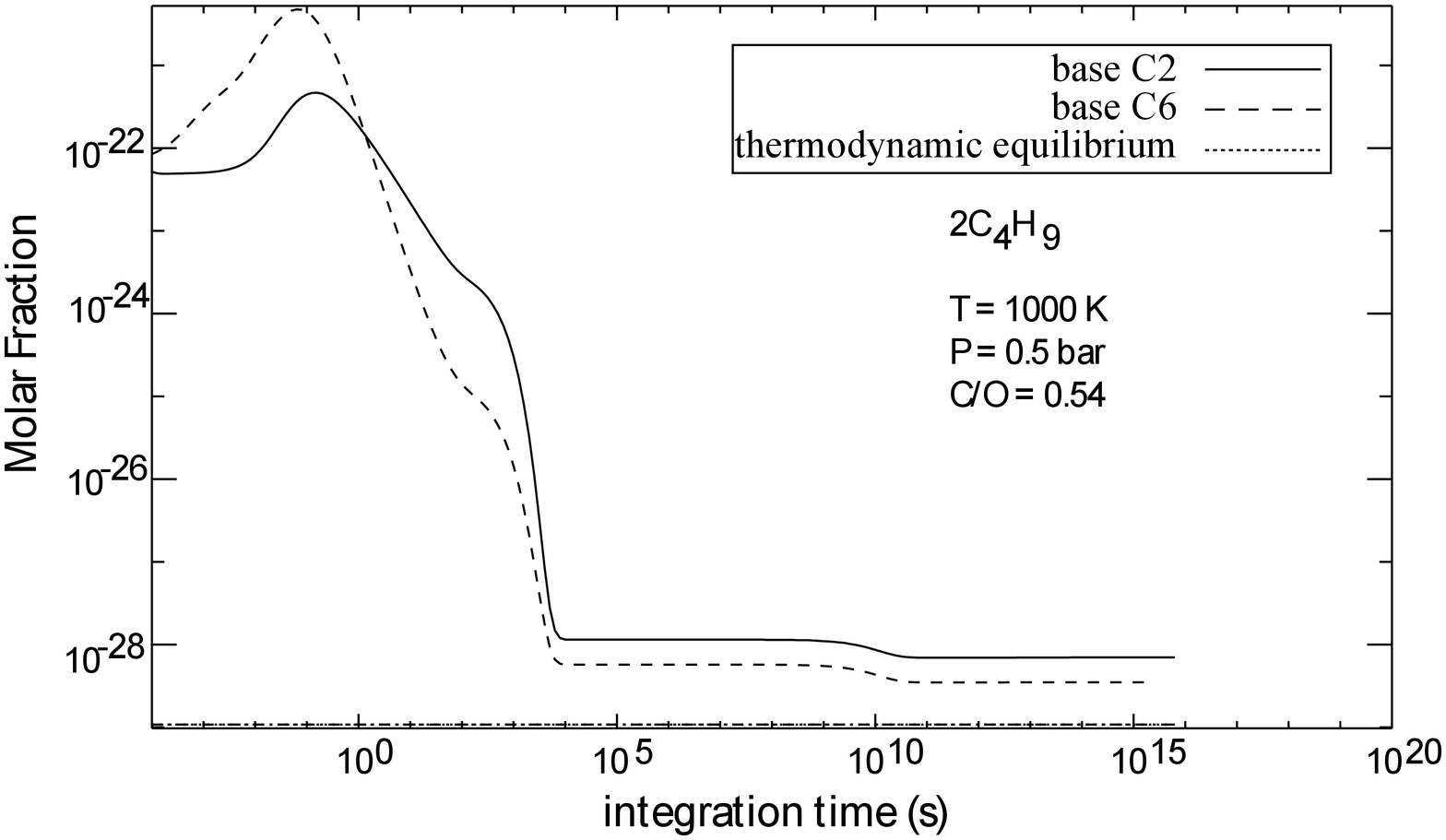}
\caption{Kinetic evolution of different species with different pressure, temperature, and C/O ratios, computed with the 0D model using the C$_0$-C$_2$ scheme (full line) or the C$_0$-C$_6$ scheme (dashed line). Kinetic evolutions reach (or evolve toward) thermochemical equilibrium (dotted line).} \label{fig:0D}
\end{figure*}
\subsection{Chemical scheme}
\subsubsection{Reaction network}

The C$_0$-C$_6$ reaction scheme includes the C$_0$-C$_2$ reaction network described in \cite{venot2012chemical}, but also reactions of C$_3$-C$_6$ unsaturated species, such as C$_3$H$_6$ (propene) or C$_4$H$_6$ (butadiene), as well as reactions of small aromatic compounds up to C$_8$H$_{10}$ (ethylbenzene).
The base of reactions includes the following sub-mechanisms:
\begin{itemize}
 \item A primary mechanism including reactions of C$_7$H$_8$ (toluene) and of C$_7$H$_7$ (benzyl), C$_6$H$_4$CH$_3$ (methylphenyl), C$_6$H$_5$CH$_2$OO (peroxybenzyl), C$_6$H$_5$CH$_2$O (alcoxy benzyl), HOC$_6$H$_4$CH$_2$O (hydroxyalcoxybenzyl), OC$_6$H$_4$CH$_3$ (cresoxy), C$_6$H$_5$CHOH, and HOC$_6$H$_4$CH$_2$ (hydroxybenzyls) free radicals. This mechanism is described in \cite{bounaceur2005experimental}.
\item A secondary mechanism involving the reactions of C$_6$H$_5$CHO (benzaldehyde), C$_6$H$_5$CH$_2$OOH (benzyl hydroperoxide), HOC$_6$H$_4$CH$_3$ (cresol), C$_6$H$_5$CH$_2$OH (benzylalcohol), C$_8$H$_{10}$ (ethylbenzene), C$_8$H$_8$ (styrene), and C$_{14}$H$_{14}$ (bibenzyl). This mechanism is also presented in \cite{bounaceur2005experimental}.
\item A mechanism for the oxidation of C$_6$H$_6$ (benzene) \citep{costa2003experimental}. It includes the reactions of C$_6$H$_6$ and of cC$_6$H$_7$ (cyclohexadienyl), cC$_6$H$_5$ (phenyl), C$_6$H$_5$O$_2$ (phenylperoxy), cC$_6$H$_5$O (phenoxy), OC$_6$H$_4$OH (hydroxyphenoxy), cC$_5$H$_5$ (cyclopentadienyl), cC$_5$H$_5$O (cyclopentadienoxy), and cC$_5$H$_4$OH (hydroxycyclopentadienyl) free radicals, as well as the reactions of C$_6$H$_4$O$_2$ (ortho-benzoquinone), cC$_6$H$_5$OH (phenol), cC$_5$H$_6$ (cyclopentadiene), cC$_5$H$_4$O (cyclopentadienone), cC$_5$H$_5$OH (cyclopentadienol), and C$_4$H$_4$O (vinylketene).
\item A mechanism for the oxidation of unsaturated C$_3$-C$_4$ species. It contains reactions involving C$_3$H$_2$ (propadienylidene),  C$_3$H$_3$ (prop-3-ynyle), the two isomers a-C$_3$H$_4$ (allene) and p-C$_3$H$_4$ (propyne), the three isomers  C$_3$H$_5$ (allyle), 1-C$_3$H$_5$ (prop-1-en-2-yle), and 2-C$_3$H$_5$ (prop-1-en-1-yle), C$_3$H$_6$ (propene), cC$_3$H$_6$ (cyclopropene), C$_4$H$_2$ (diacetylene), the two isomers n-C$_4$H$_3$ (but-1-en-3-ynyle) and i-C$_4$H$_3$ (but-1-en-3-yn-2-yle), C$_4$H$_4$ (vinylacetylene), the five isomers n-C$_4$H$_5$ (1,3-butadienyle), i-C$_4$H$_5$ (1,3-butadien-2-yle), 13-C$_4$H$_5$ ( but-1-yn-3-yle), 14-C$_4$H$_5$ (but-1-yn-4-yle), and 21-C$_4$H$_5$ (but-2-yn-1-yl), the five isomers 13-C$_4$H$_6$ (1,3-butadiene), 12-C$_4$H$_6$ (1,2- butadiene), cC$_4$H$_6$ (methyl-cyclopropene), 1-C$_4$H$_6$  (1-butyne), and 2-C$_4$H$_6$ (2-butyne). This sub-mechanism integrates reactions involved in the formation of aromatic compounds and has been developed and validated against experimental data from the literature \citep{westmoreland1989, tsang1991chemical, miller1992kinetic, lindstedt1996detailed, hidaka1996shock, wang1997detailed}.
\end{itemize}
This database has been used with success in modelling of premixed flame of butadiene, propyne, allene, and acetylene \citep{fournet1999experimental} and in predicting formation of small aromatic compounds for flame of methane \citep{gueniche2009comparative, belmekki2002experimental}.\\
To obtain a good chemical overlap between the reactions bases C$_0$-C$_2$ and C$_3$-C$_6$, we added a detailed kinetic mechanism for the oxidation of a propane/n-butane mixture. This sub-mechanism includes a comprehensive primary mechanism, where the only molecular reactants considered are the initial organic compounds (here propane and n-butane) and oxygen, and a lumped secondary mechanism that contains the reactions consuming the molecular products of the primary mechanism that do not react in the reaction bases C$_0$-C$_2$ or C$_3$-C$_6$.
This sub-mechanism has been used with success in modelling laminar flame velocity for components of natural gas \citep{dirrenberger2011measurements}.

In this reaction base, we included pressure-dependent rate constants for unimolecular decomposition, recombination, beta-scission, and additional reactions following the formalism proposed by \cite{troe1974}, as well as third-body efficiency coefficients.
As for the C$_0$-C$_2$ chemical scheme, all the reactions of the C$_0$-C$_6$ scheme are reversed, which allows reproducing thermochemical equilibrium. Most of them (called "reversible reactions") are reversed through the principle of microscopic reversibility (e.g. $K_{eq} = k_f/k_r$, with $K_{eq}$ the equilibrium constant, $k_f$ the reaction rate of the forward reaction, and $k_r$ the reaction rate of the reverse reaction, see \citealt{venot2012chemical} for more details), but a few of them (called "irreversible reactions") are reversed using experimentally measured reaction rates for both the forward and the reverse reaction. This ensures a better reproduction of the out-of-equilibrium experiments.

Thermochemical data for molecules or radicals were calculated and stored as 14 polynomial coefficients according to the CHEMKIN formalism \citep{chemkin1996}. These data were calculated using the software THERGAS \citep{muller1995}, which is based on the group and bond additivity methods proposed by \cite{Benson76}.

In summary, the C$_0$-C$_6$ reaction base is composed of 240 reactants that are involved in 4002 chemical reactions: 1991 reversible reactions and 20 irreversible reactions. This chemical scheme is available at the online database KIDA: Kinetic Database for Astrochemistry\footnote{http://kida.obs.u-bordeaux1.fr/models/}\citep{KIDA2012}.

\subsubsection{Photolysis}
To model the out-of-equilibrium process that is due to photolysis, we added a set of 113 photodissociations to the C$_0$-C$_6$ chemical scheme. We modelled the stellar irradiation by using the solar spectrum \citep{thuillier2004solar} in the range [0-900] nm.\\
For most molecules, no data at high temperature exist, so we used absorption cross-sections at ambient temperature that we found in the MPI-Mainz UV/VIS Spectral Atlas \citep{MPIdatabase2013} and in the UV/Vis+ Spectra Data Base\footnote{http://www.uv-spectra.de/}. More specific references and explanations of the methodology of calculating cross-sections and quantum yields can be found in \cite{venot2012chemical}, \cite{2013Hebrard}, and \cite{dob2014}. For CO$_2$ \citep{venot2013high} and NH$_3$ \citep{venot_NH3}, we used our recent measurements at 500~K.
As we present in Sect.~\ref{sec:1Dmodel}, we used three different thermal profiles. For all of them, we used the same cross-sections at 500~K and did not try to adjust the cross-section of CO$_2$ to the exact temperature at which photodissociations occurred because we are mainly interested in the effect of the new chemical scheme at different atmospheric temperatures. However, we verified with the two different models ($T_{1000}$~$\zeta_{0.54}$ and $T_{1000}$~$\zeta_{1.1}$, see Sect.~\ref{sec:1Dmodel} and Table~\ref{table:models_grid} for more explanations and the meaning of these symbols) that this approximation has no significant effect on the computed chemical abundances. The differences are always lower than 10\%, and for CO$_2$, even lower than 0.54\% for solar C/O. When C/O=1.1, the differences are lower than 30\% and lower than 6.7\% for CO$_2$. These differences would not be visible on the abundances profiles figures plotted with a logarithmic scale.

\subsection{0D models}\label{sec:0D}

First we compared the kinetic evolution of species calculated with the two chemical schemes.
We used the 0D model presented in \cite{venot2012chemical} that allows determining the chemical evolution of a mixture at a certain pressure and temperature, with no photolysis process. We computed several conditions corresponding to the combination of $T =$ 500, 800, 1000, or 1500~K, $P =$ 0.5 or 50~bar, and C/O=0.54 or 1.1, with the C$_0$-C$_2$ and the C$_0$-C$_6$ scheme. Initial conditions correspond to the thermochemical equilibrium of the bottom level at the profile $T_{500}$, that is, $P$=1000~bar and $T$=1742~K with C/O=0.54 or 1.1.\\
For most species and regardless of the $P$, $T$, and C/O conditions, the kinetic evolution predicted by the two chemical schemes is the same and reaches thermochemical equilibrium. This is demonstrated for CO$_2$ in the first panel of Fig.~\ref{fig:0D}. Nevertheless, for some hydrocarbon species, we observe a slight difference in the kinetic evolution and sometimes also in the steady state, as can be seen in Fig.~\ref{fig:0D}. These differences are found for species with more than two carbon atoms, which is normal because the C$_0$-C$_2$ scheme is not made to study these heavy species. It contains C$_n$H$_x$ species, with $n>2$, only to ensure that species with two or fewer carbon atoms will have a correct behaviour. The abundances and the kinetic evolutions of these heavy species found with the C$_0$-C$_2$ scheme must not be trusted. Thus, it is expected that we find a difference in evolution for some of these hydrocarbons.

\subsection{Grid of 1D models}\label{sec:1Dmodel}

To construct the thermal profiles, we used the analytical model of \cite{parmentier2014a}\footnote{https://www-n.oca.eu/parmentier/nongrey/nongrey.html} that was calibrated to match numerical $P$-$T$ profiles of solar-composition clear-sky atmospheres by \cite{Parmentier2015}. We used the coefficients from \cite{Parmentier2015} and the opacities from \cite{valencia2013bulk}. For simplicity, we did not consider the presence of TiO, which may cause a thermal inversion in hot atmospheres. Moreover, it is still unclear whether there is thermal inversion in hot Jupiters (and whether this is caused by TiO) \citep[e.g.][]{Madhu2014, Parmentier2014_expastro}. As an illustration, \cite{diamond2014} reanalysed the Spitzer data of HD~209458b and found no evidence for a stratosphere, which contradicts previous studies claiming that the atmosphere of this exoplanet presented a thermal inversion \citep{burrows2007,Knutson2008}.

\begin{figure}[!hb]
\centering
\includegraphics[angle=0,width=\columnwidth]{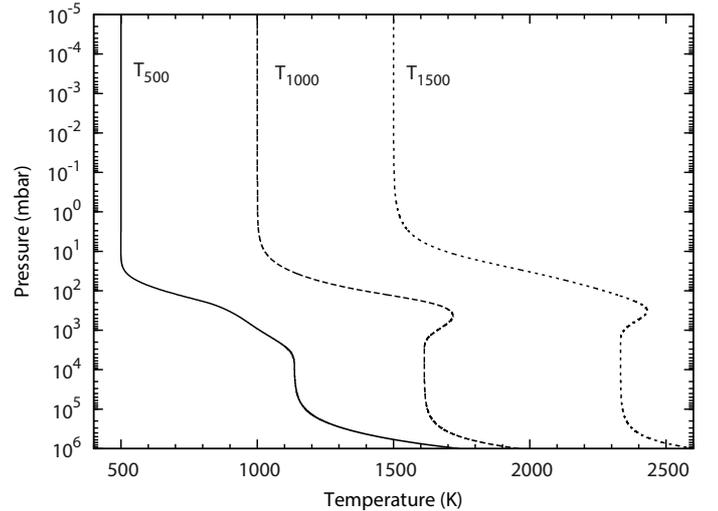}
\caption{Vertical profiles of temperature. The cool atmosphere ($T_{500}$) has an isothermal part from $\sim$20 to 10$^{-5}$ mbar at 500~K(full line). The warm atmosphere ($T_{1000}$) has an isothermal part from $\sim$2 to 10$^{-5}$ mbar at 1000~K (dashed line). Finally, the hot atmosphere ($T_{1500}$) has an isothermal part from $\sim$0.3 to 10$^{-5}$ mbar at 1500~K (dotted line).} \label{fig:profile_pT}
\end{figure}

To span a range of conditions relevant to known close-in giant exoplanets, we selected as baselines of our study three thermal profiles with high-altitude atmospheric temperatures of 500~K, 1000~K, and 1500~K. These temperature-pressure profiles are shown in Fig.~\ref{fig:profile_pT}. To obtain these profiles, we set the irradiation temperature, $T_{irr}$, to 784, 1522, and 2303~K. We considered that $\mu$ = 1/$\sqrt{3}$, where $\mu$ = $\cos \theta$ and $\theta$ the inclination of the stellar irradiation with respect to the local vertical direction. This choice allowed us to obtain dayside average profiles. We considered that the planets have a low internal temperature (T$_{int}$~=~100~K) and a gravity of 25 m.s$^{-2}$. At the time of the submission of our study, a new version of the analytical model was released. In this version, the user can no longer directly choose the irradiation temperature, but can set the equilibrium temperature, $T_{eq0}$, as well as a parameter $f$, which modulates the flux received by the planet. The profiles used in this study correspond approximately to dayside average profiles ($f$ = 0.5) of planets with an equilibrium temperature for zero albedo of $T_{eq0}$ = 627, 1122, and 1718~K for the cool, warm, and hot profiles.
Since the goal of this study is to compare the C$_0$-C$_2$ and the C$_0$-C$_6$ chemical schemes in different conditions, we used the same thermal profiles when considering a solar and not solar C/O ratio, even if the profiles were calculated assuming a solar composition. The C/O ratio varies between two values: 0.54 (solar) and 1.1. Finally, we compared the results including or excluding photolysis processes. The different parameters used in the models are summarised in Table~\ref{table:models_grid} with the corresponding symbols. The planet-star distance of each planet has been adjusted to match their equilibrium temperature. For the cool, warm, and hot profiles, they have been set to $d$=0.2196, 0.0549, and 0.0244 AU, respectively. Because many uncertainties exist on the vertical mixing acting in exoplanet atmospheres, we used a constant eddy diffusion coefficient, $K_{zz}$=10$^8$ cm$^2$.s${-1}$, and did not try to adjust a complex eddy diffusion profile. This value is similar to what has been used or calculated by previous models \citep[i.e.][]{lew2010, moses2011disequilibrium, line2011thermochemical, venot2013high}. However, we recall that this value might be too high as pointed out by \cite{par2013}. We did not explore the space of possible values for the eddy diffusion coefficient because it was beyond the scope of this paper. Nevertheless, with a stronger vertical mixing, quenching would occur in deeper layers, and for the $T_{1000}$ and $T_{1500}$ profiles, the CO/CH$_{4}$ ratio could become lower than unity in the entire atmosphere \citep{venot2014GJ3470b}.

\begin{table}
\caption{Grid of parameters.} \label{table:models_grid}
\centering
\begin{tabular}{llll}
\hline \hline
Parameter      & Range of values                                         & Symbol \\
\hline
Temperature & Cool atmosphere (500~K)        & $T_{500}$ \\
                     & Warm atmosphere (1000~K)    & $T_{1000}$ \\
                     & Hot atmosphere (1500~K)        & $T_{1500}$\\
\hline
C/O ratio     & 0.54 (solar)       & $\zeta_{0.54}$ \\
                   & 1.1  & $\zeta_{1.1}$ \\
\hline
Stellar UV flux                   & Irradiation  & $F_{On}$ \\
                                         &  No irradiation   & $F_{Off}$ \\
\hline
\end{tabular}
\end{table}

\section{Results}\label{sec:Resultats}

As can be seen in Figs.~\ref{fig:sanshv} and \ref{fig:avechv}, the chemical composition of the atmospheres varies depending on the temperature and the C/O ratio. The photodissociations and the chemical scheme used also have an influence. In the following, we comment on these differences and explain them for some species. Indeed, by inspecting the reaction rates, we have identified some formation pathways that occur only in the C$_0$-C$_6$ chemical scheme and that can explain the departures between the two reactions networks. For species for which the two chemical schemes predict different mixing ratios, we determined from which reactions these compounds were mainly formed. We compared their reaction rates in the two chemical schemes to see which reaction was more efficient in one of the schemes. Then we examined through which reaction the reactants of this latter reaction were formed. We proceeded iteratively until we identified a complete formation pathway. This identification is a preliminary work. Determining the detailed production and destruction pathways of species with precision requires the use of algorithms such as the \textit{Pathway Analysis Program} \citep{lehmann2004algorithm, stock2012chemical}. These algorithms are very powerful, but need to be adapted to chemical schemes as large as our C$_0$-C$_6$ chemical network.
For the clarity of Figs.~\ref{fig:sanshv} and \ref{fig:avechv}, we chose to plot only species with significant abundances ($>$ 10$^{-7}$), but numerous carbon species have an abundance between 10$^{-10}$ and 10$^{-7}$. We have indicated these species in Table~\ref{table:abundances_grid}.

\subsection{Chemical equilibrium}

At chemical equilibrium there are some differences between the different thermal profiles and C/O ratios (see the dotted lines in Fig.~\ref{fig:sanshv}).
Regardless of the C/O ratio, the hotter the atmosphere, the higher the CO/CH$_4$ abundance ratio. CH$_4$ becomes more abundant than CO at low temperatures, while CO dominates CH$_4$ at high temperatures. Moreover, the abundance of H$_2$O decreases when the temperature increases.
We remark that when the C/O ratio increases to above unity, then the abundances of carbon-bearing species such as CH$_4$, CH$_3$, and HCN increase their abundances significantly. These results agree with previous studies on the effect of the C/O ratio on the chemical atmospheric composition \citep[i.e.][]{Madhu2011carbonrich, Madhu2012, moses2013chemical}.
When H$_2$O is very abundant, OH behaves similarly to H, but with a smaller amount. However, when the abundance of H$_2$O is quite low ($\zeta_{1.1}$ for $T_{1000}$ and $T_{1500}$ profiles), the amount of OH is limited by that of water, and OH behaves like H$_2$O. Indeed, in each case tested here, the main reaction that produces OH is H + H$_2$O $\rightarrow$ OH + H$_2$, which means that OH is limited by the reactant of the less abundant species.

\subsection{Without photodissociation}

\begin{figure*}
\centering
\includegraphics[angle=0,width=\columnwidth]{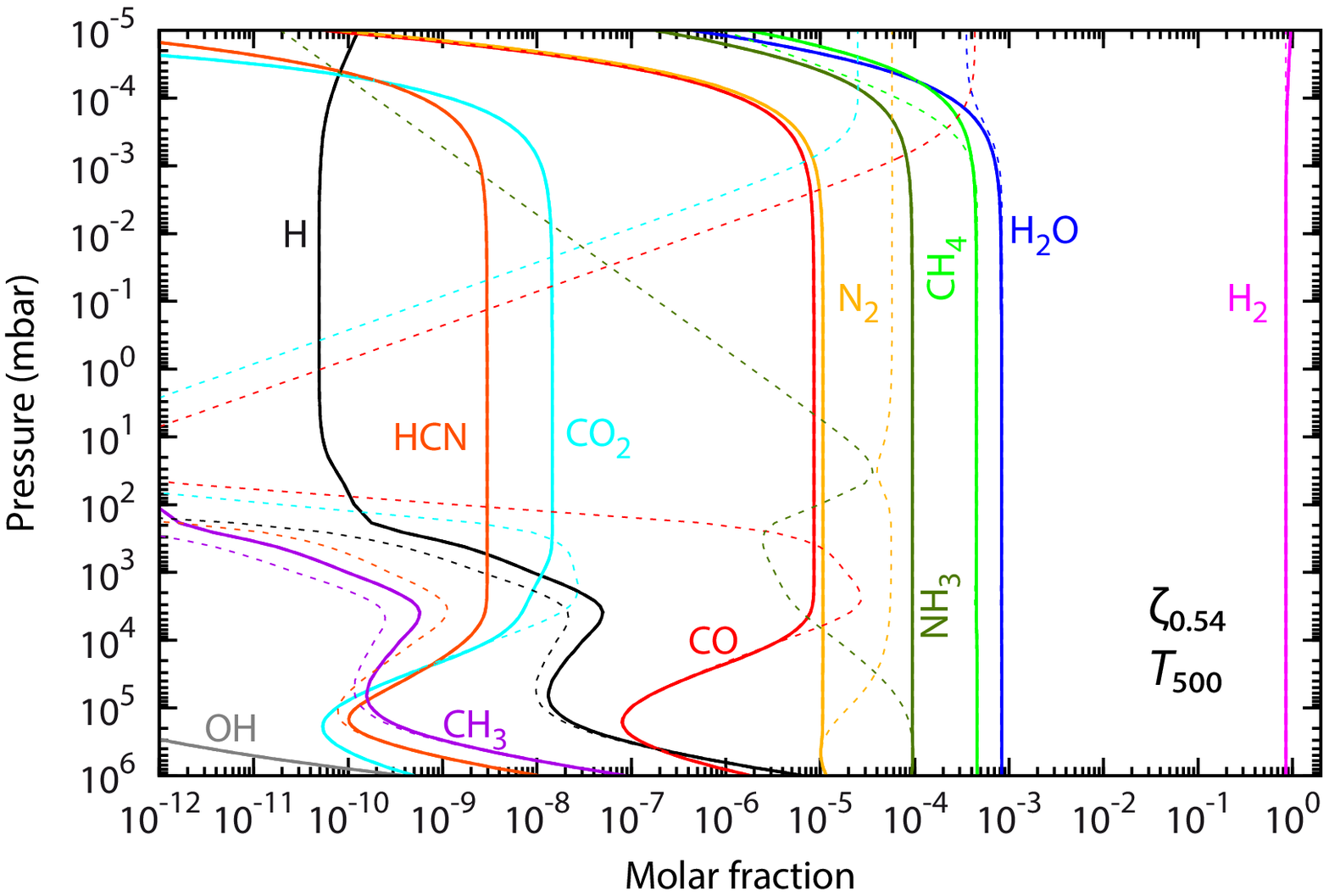}
\includegraphics[angle=0,width=\columnwidth]{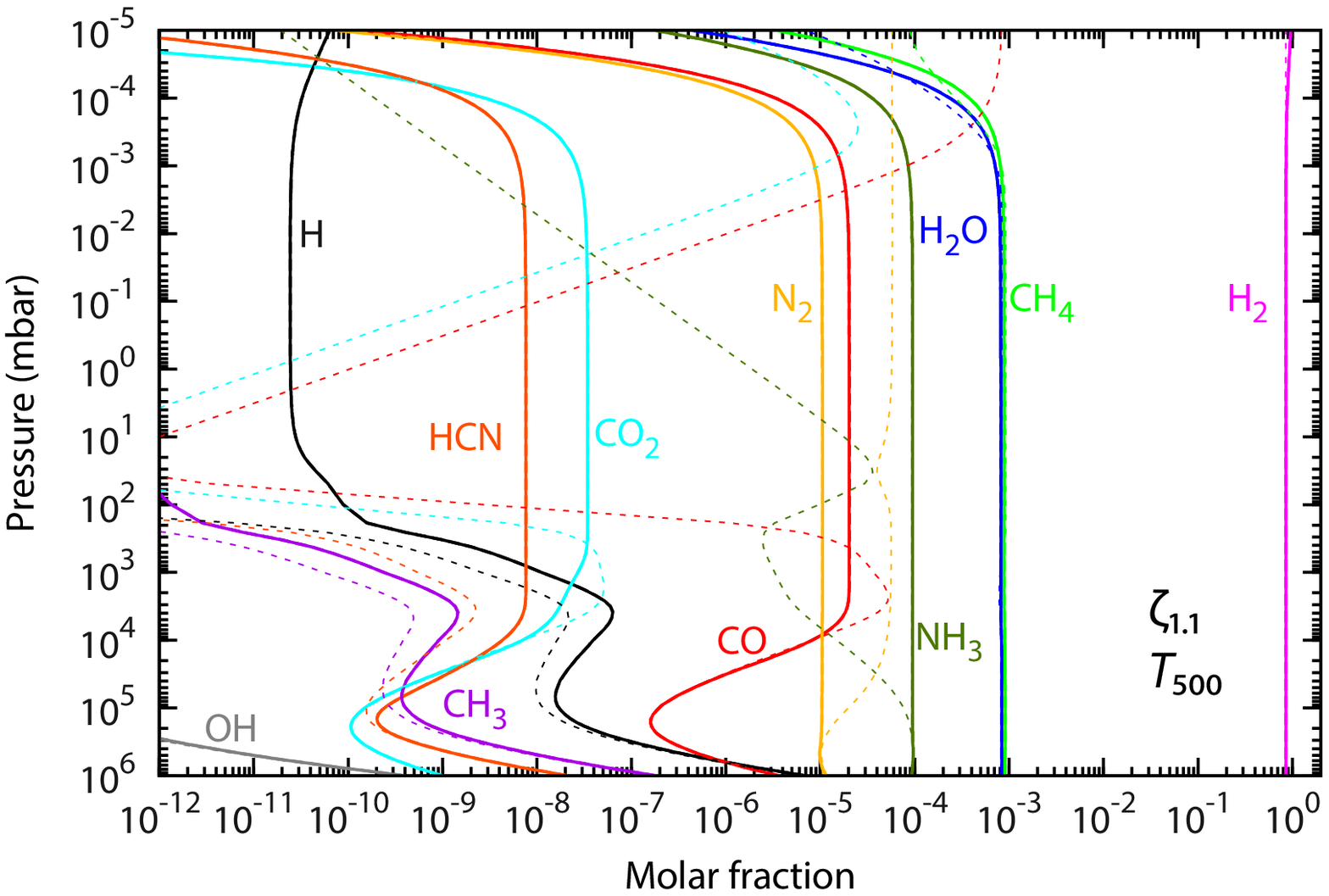}
\includegraphics[angle=0,width=\columnwidth]{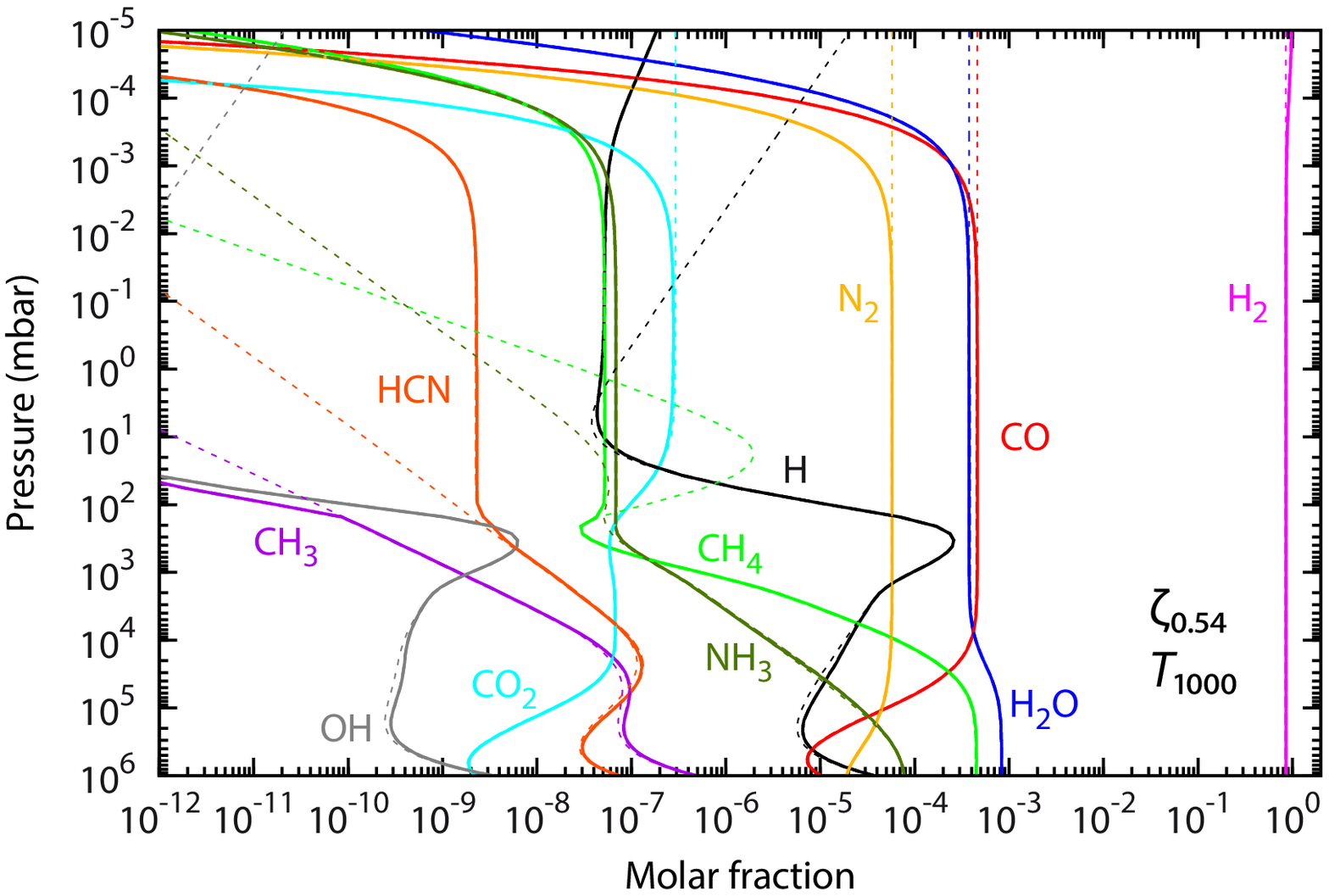}
\includegraphics[angle=0,width=\columnwidth]{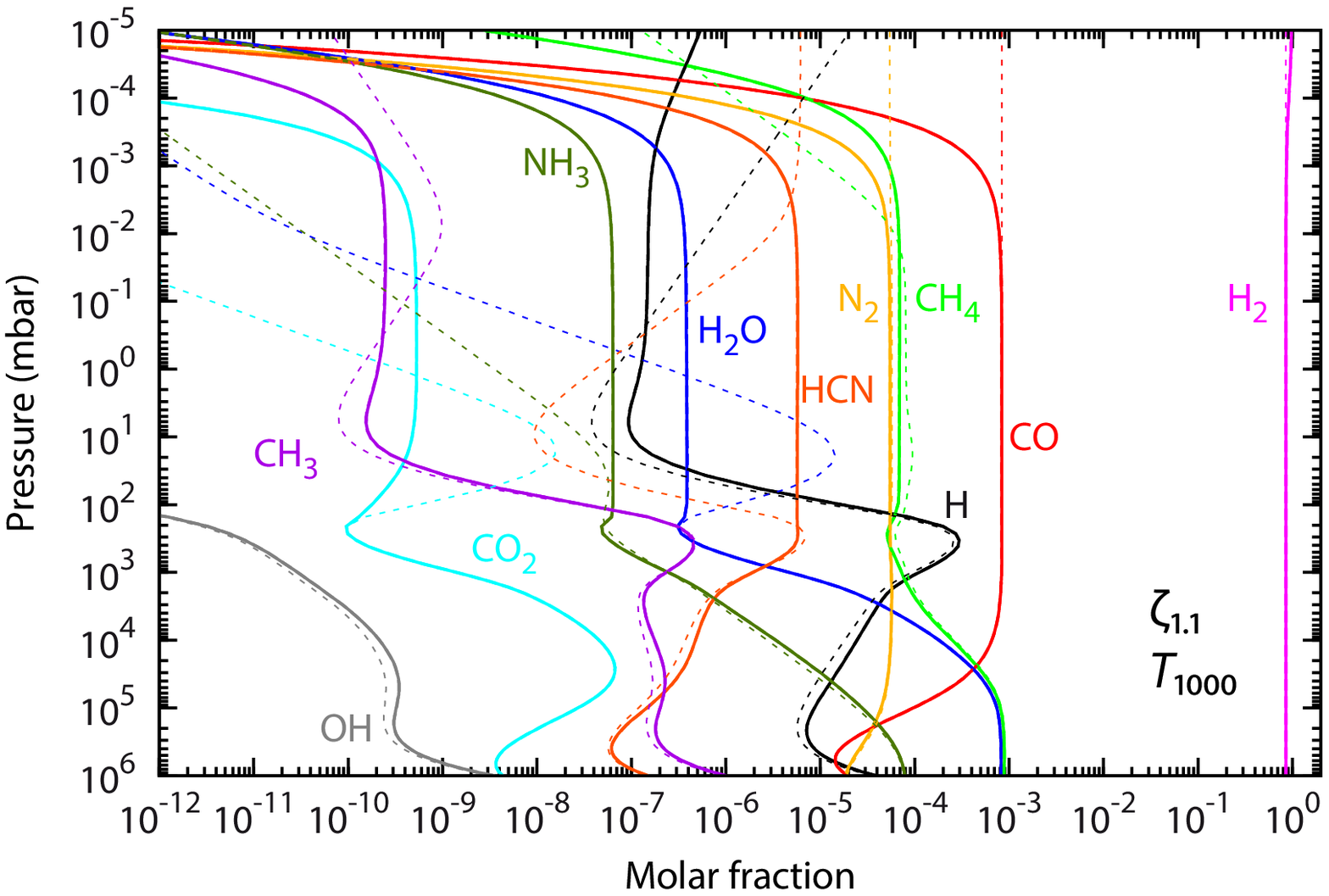}
\includegraphics[angle=0,width=\columnwidth]{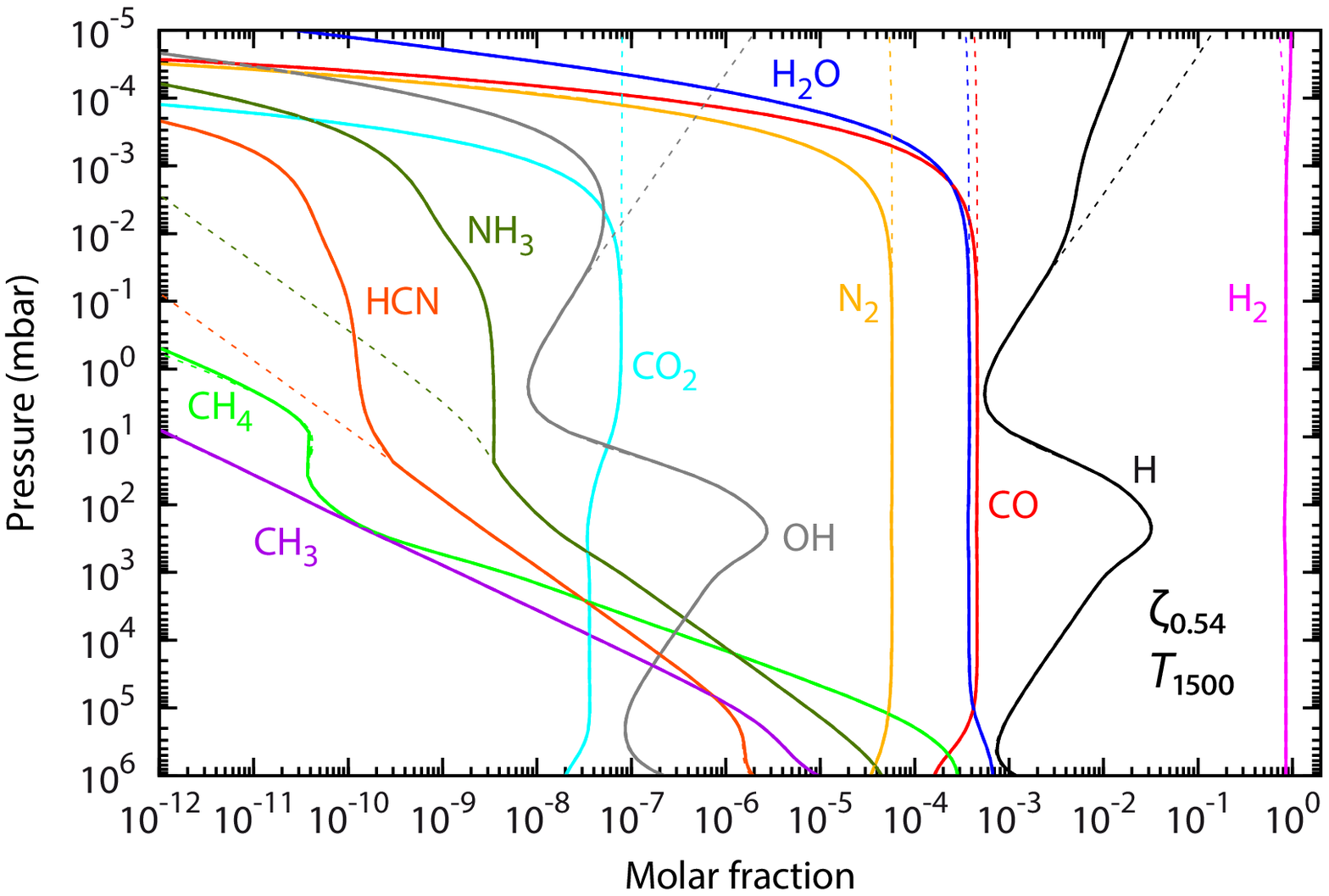}
\includegraphics[angle=0,width=\columnwidth]{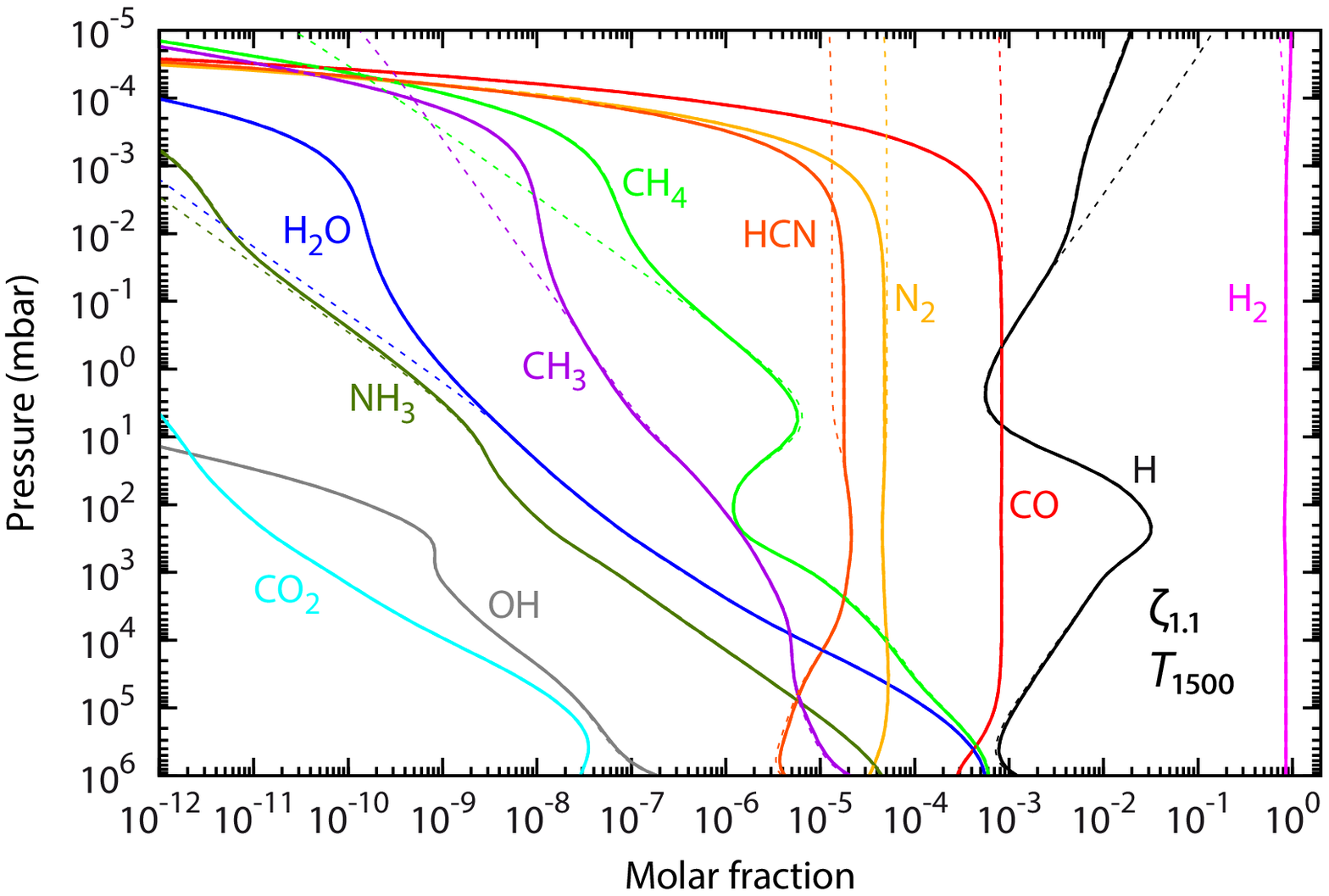}
\caption{Vertical abundance profiles for the cool (up), warm (middle), and hot (bottom) thermal profiles, with the C/O ratio solar (left) and C/O=1.1 (right). The chemical compositions are calculated with the C$_0$-C$_2$ scheme (dashed line) and the C$_0$-C$_6$ scheme (solid line), without photodissociation. Dashed and solid lines overlap, so that the dashed lines are obscured. The chemical equilibrium (dotted line) is also represented.} \label{fig:sanshv}
\end{figure*}

The chemical compositions given by the C$_0$-C$_2$ and the C$_0$-C$_6$ chemical schemes are identical (considering species that are present in both networks) for all thermal profiles and the two C/O ratios when photodissociation is omitted. The only departure between the two schemes concerns some C$_{n>2}$H$_x$ species. As we explained in Sect.~\ref{sec:0D}, the C$_0$-C$_2$ contains species with three or four carbon atoms to correctly describe the kinetics of the species with two carbon atoms, but the kinetic behaviour of these heavier compounds must not be trusted. Thus, it is expected that the C$_0$-C$_6$ scheme predicts a different behaviour for these hydrocarbons. Nevertheless, these small departures do not affect the abundances of the main species represented in Fig.~\ref{fig:sanshv}.

The quenching level is different depending on the thermal profile. Logically, the cooler the profile, the deeper quenching occurs in the atmosphere. For $T_{500}$, quenching occurs between 10$^5$ and 10$^4$ mbar, depending on the molecule, whereas for $T_{1500}$, it occurs much higher in the atmosphere, at about 20 mbar. 

\subsection{With photodissociation}

\begin{figure*}
\centering
\includegraphics[angle=0,width=\columnwidth]{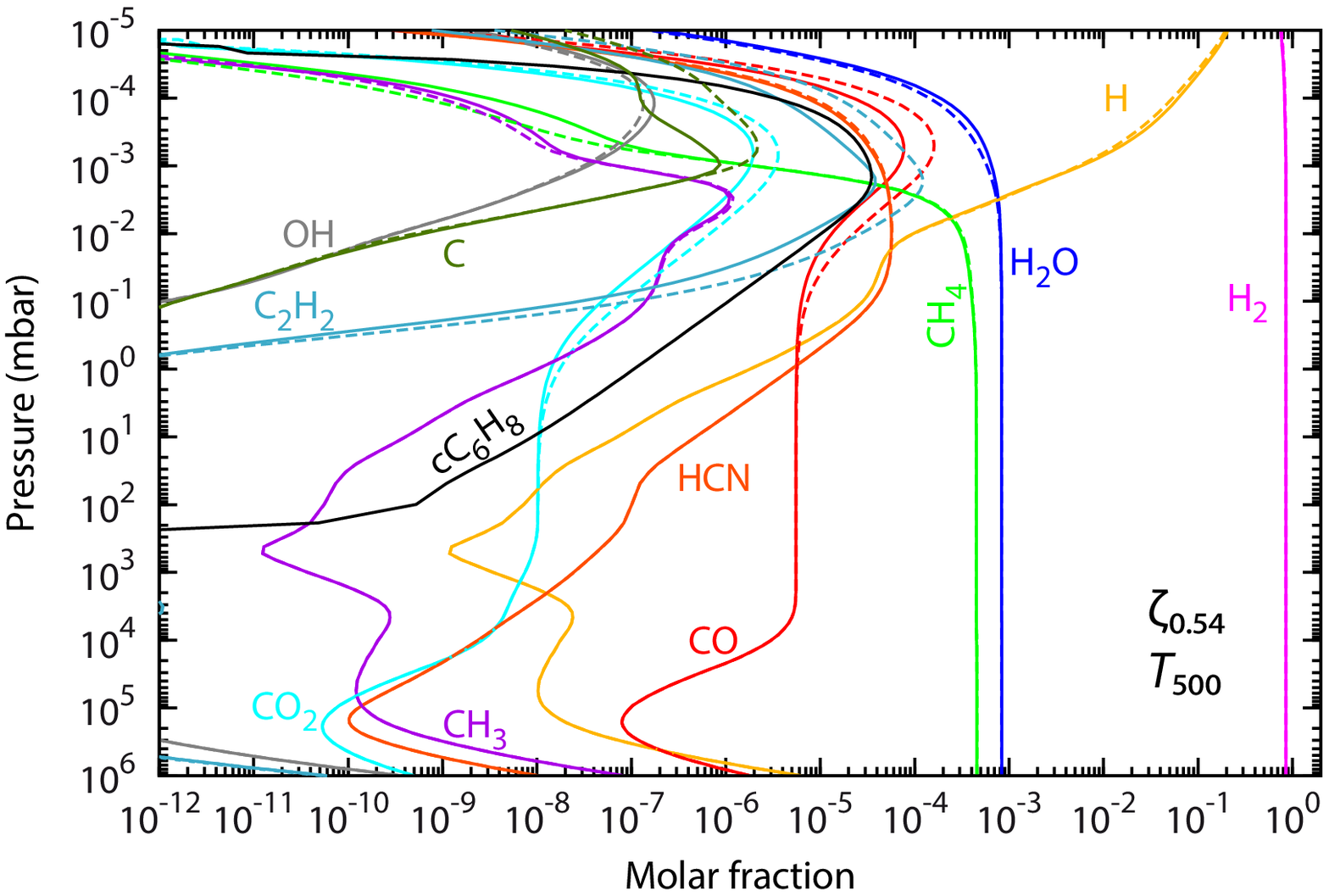}
\includegraphics[angle=0,width=\columnwidth]{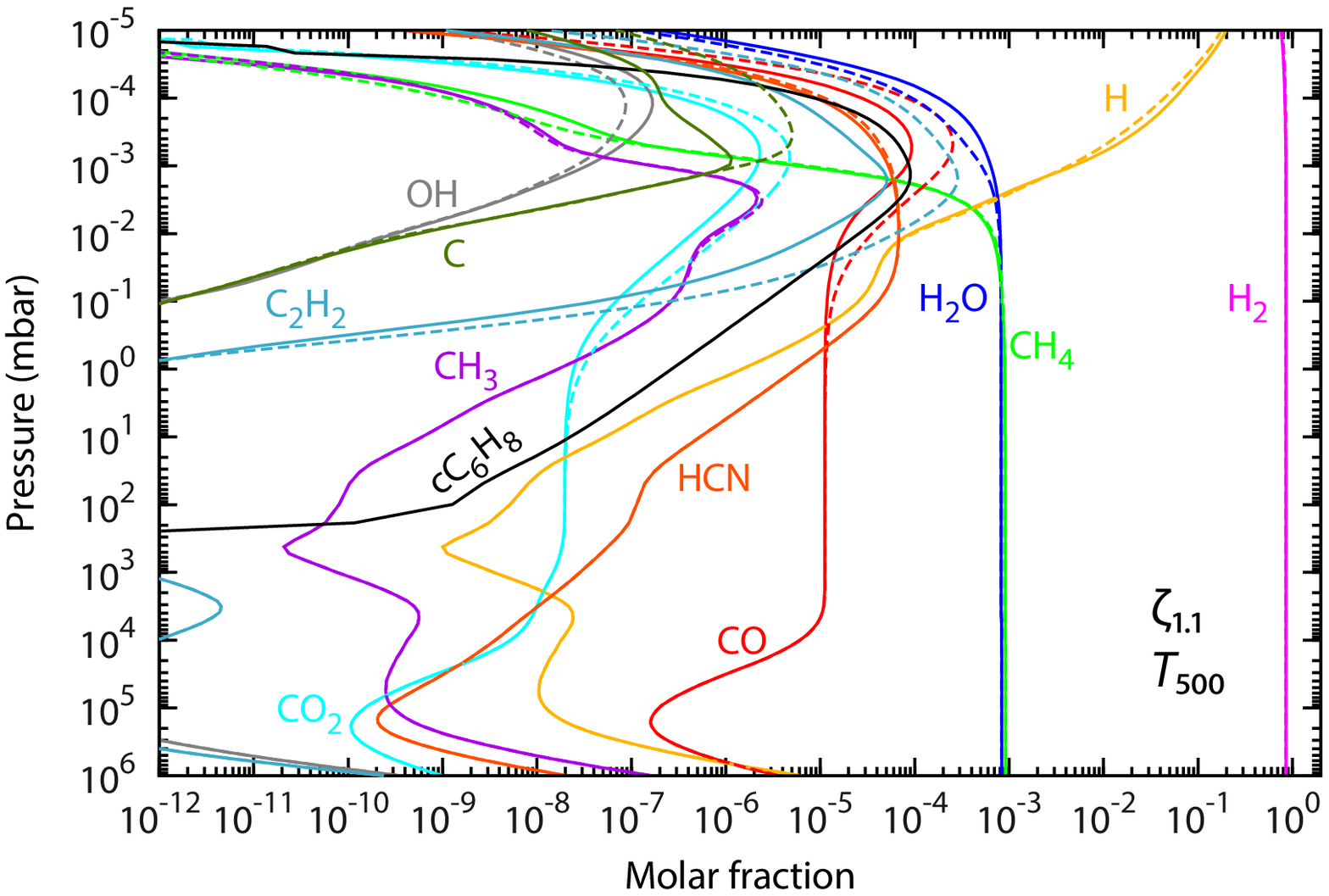}
\includegraphics[angle=0,width=\columnwidth]{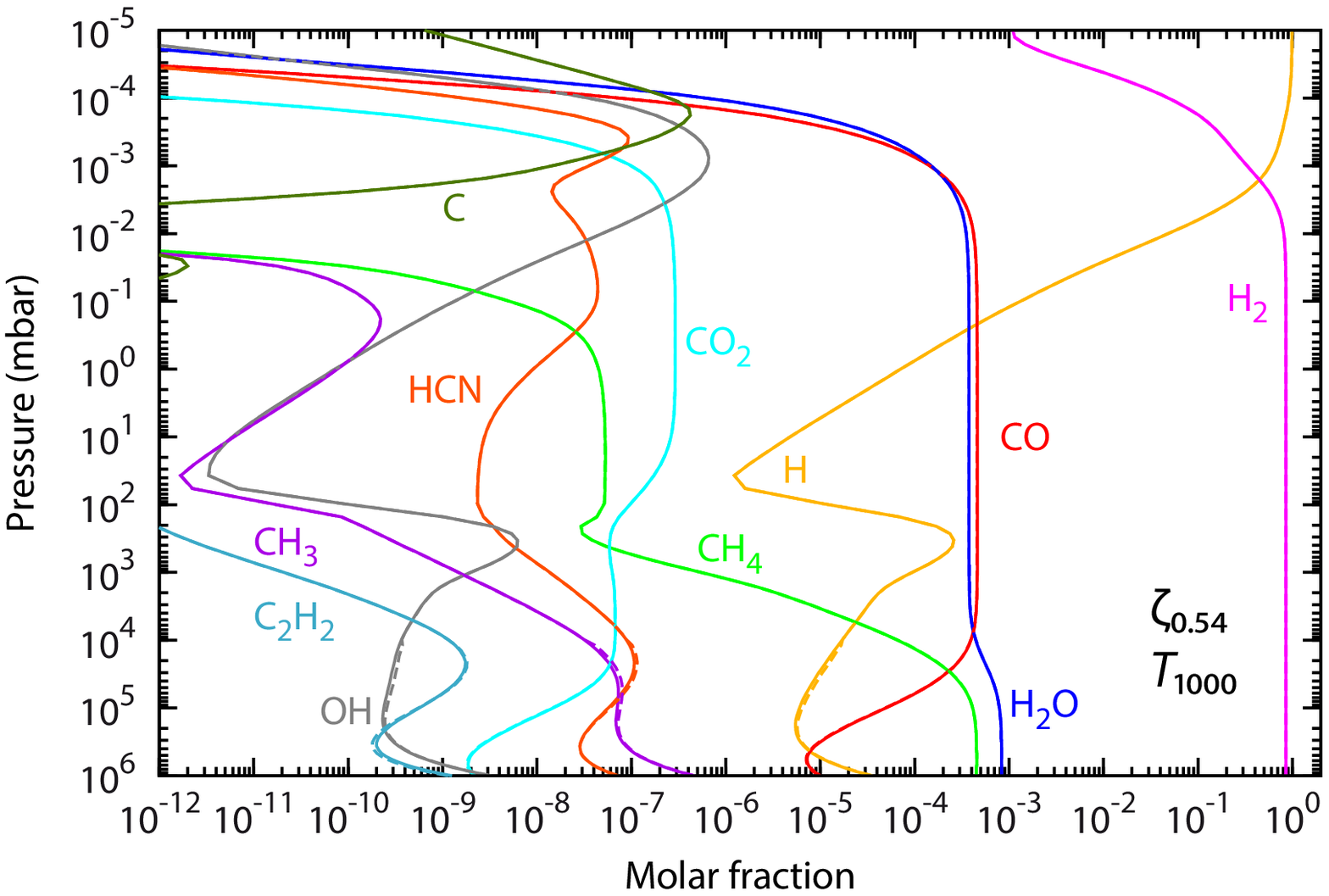}
\includegraphics[angle=0,width=\columnwidth]{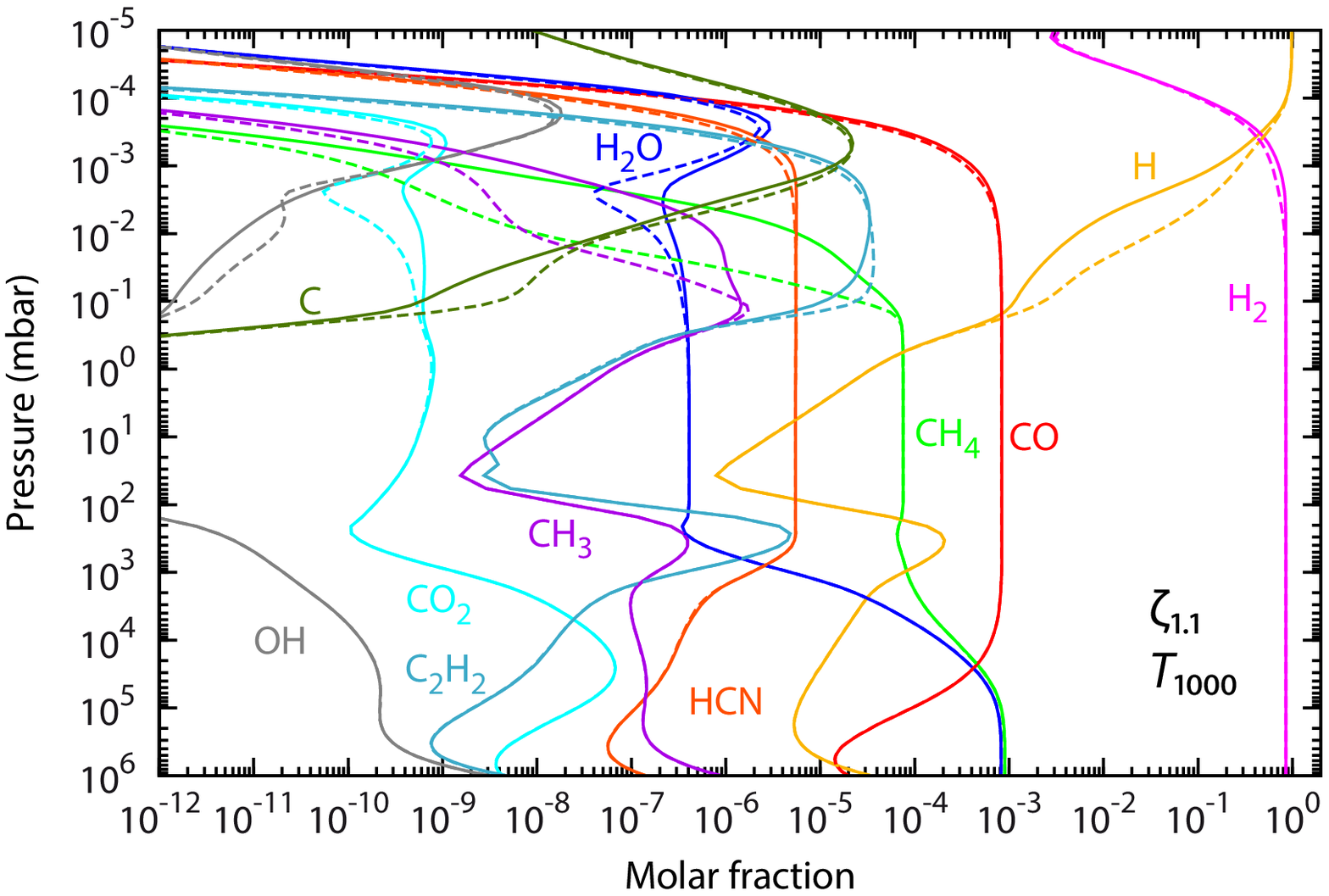}
\includegraphics[angle=0,width=\columnwidth]{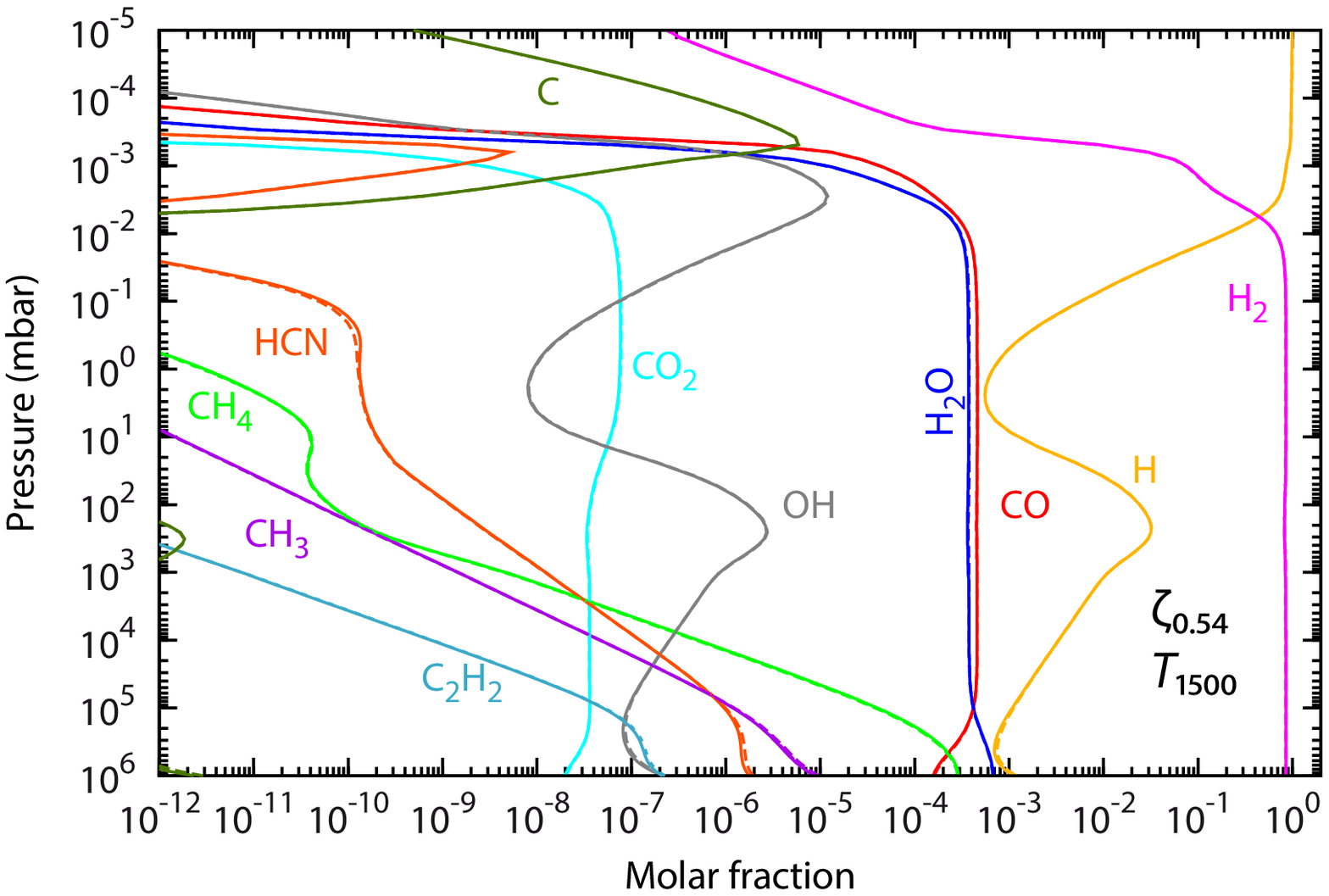}
\includegraphics[angle=0,width=\columnwidth]{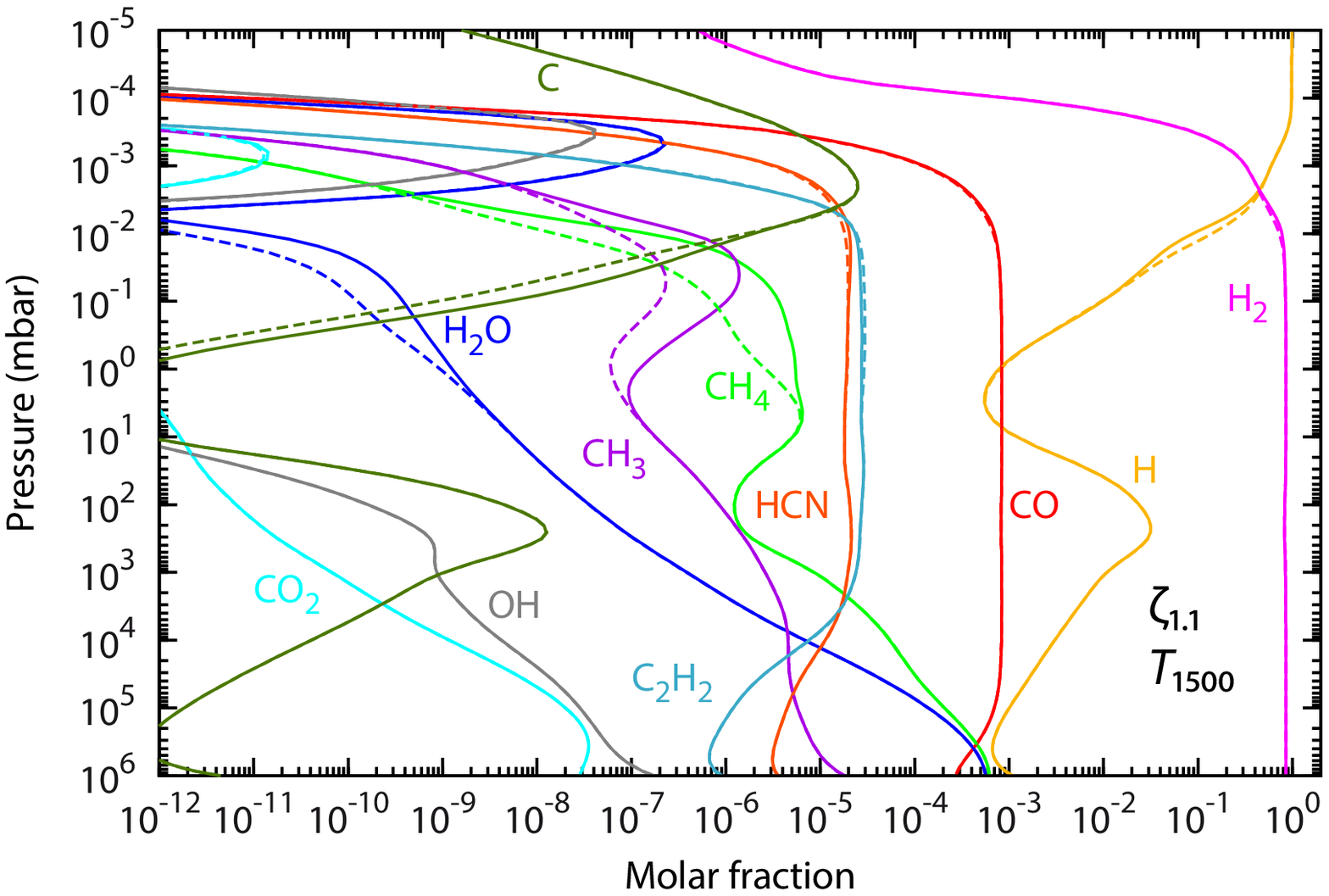}
\caption{Vertical abundance profiles for the cool (up), warm (middle), and hot (bottom) thermal profiles, with the C/O ratio solar (left) and C/O=1.1 (right). The chemical compositions are calculated with the C$_0$-C$_2$ scheme (dashed line) and the C$_0$-C$_6$ scheme (solid line), with photodissociation.} \label{fig:avechv}
\end{figure*}

\begin{table*}
\caption{Highest abundances of some hydrocarbon species not plotted in Fig.\ref{fig:avechv}, but with an abundance higher than 10$^{-10}$. These mixing ratios are obtained using the C$_0$-C$_6$ scheme with the six models including photodissociations. The level pressure (in mbar) at which these abundances are reached is noted between parenthesis. Hydrocarbons that are not indicated in this table have an abundance always lower than 10$^{-10}$.} \label{table:abundances_grid}
\centering
\begin{tabular}{l|cccccccc}
\hline \hline 
Species      & $T_{1500}$  $\zeta_{0.54}$        & $T_{1500}$  $\zeta_{1.1}$ & $T_{1000}$  $\zeta_{0.54}$        & $T_{1000}$  $\zeta_{1.1}$ & $T_{500}$  $\zeta_{0.54}$        & $T_{500}$  $\zeta_{1.1}$\\
\hline 

CH  &    $<$   10$^{-10}$     &  4.4$\times$10$^{-8}$ (3$\times$10$^{-3}$) &    $<$   10$^{-10}$  & 2.0$\times$10$^{-9}$ (9$\times$10$^{-4}$) & $<$   10$^{-10}$  &  1.2$\times$10$^{-10}$ (2$\times$10$^{-3}$)\\

$^1$CH$_2$    &  1.7$\times$10$^{-9}$ (10$^6$)     &  7.0$\times$10$^{-9}$ (3$\times$10$^{2}$) & $<$   10$^{-10}$  & $<$   10$^{-10}$  & $<$   10$^{-10}$  & $<$   10$^{-10}$\\

$^3$CH$_2$    &  2.4$\times$10$^{-8}$ (10$^6$)     &  1.1$\times$10$^{-7}$ (3$\times$10$^{2}$) & $<$   10$^{-10}$  &   8.3$\times$10$^{-9}$ (8$\times$10$^{-3}$) & 2.1$\times$10$^{-9}$ (3$\times$10$^{-3}$) & 4.6$\times$10$^{-9}$ (3$\times$10$^{-3}$) \\

HCO  &    2.5$\times$10$^{-8}$ (10$^6$)     &  4.3$\times$10$^{-8}$ (10$^6$) &    1.8$\times$10$^{-10}$ (10$^6$)     & 3.4$\times$10$^{-10}$ (10$^6$)     &  $<$   10$^{-10}$  & $<$   10$^{-10}$\\

H$_2$CO  &    7.9$\times$10$^{-8}$ (10$^6$)     &  1.4$\times$10$^{-7}$ (10$^6$) &    5.7$\times$10$^{-9}$ (10$^6$)     & 1.1$\times$10$^{-8}$ (10$^6$)     & 1.1$\times$10$^{-9}$ (2$\times$10$^{-3}$)     & 2.8$\times$10$^{-9}$ (2$\times$10$^{-3}$) \\

CH$_2$OH    &  1.2$\times$10$^{-10}$ (10$^6$)     &  2.0$\times$10$^{-10}$ (10$^6$) & $<$   10$^{-10}$  & $<$   10$^{-10}$  & $<$   10$^{-10}$\\

CH$_3$OH    &  1.6$\times$10$^{-9}$ (10$^6$)     &  2.7$\times$10$^{-10}$ (10$^6$) &  4.5$\times$10$^{-10}$ (10$^6$)     & 8.9$\times$10$^{-10}$ (10$^6$)     & 1.8$\times$10$^{-10}$ (10$^6$)     & 3.6$\times$10$^{-10}$ (10$^6$)   \\

C$_2$H  &    $<$   10$^{-10}$  &  1.6$\times$10$^{-7}$ (2$\times$10$^{2}$)  &    $<$   10$^{-10}$  & 5.0$\times$10$^{-9}$ (9$\times$10$^{-4}$)  &  $<$   10$^{-10}$  & $<$   10$^{-10}$\\

CH$_2$CO  &    1.9$\times$10$^{-10}$ (10$^6$)     &  6.8$\times$10$^{-10}$ (10$^6$) & $<$   10$^{-10}$  & $<$   10$^{-10}$  & $<$   10$^{-10}$  & $<$   10$^{-10}$\\

CH$_2$CHO &    $<$   10$^{-10}$  &    $<$   10$^{-10}$  &    $<$   10$^{-10}$  &  $<$   10$^{-10}$  & 6.5$\times$10$^{-10}$ (2$\times$10$^{-3}$) & 1.6$\times$10$^{-9}$ (2$\times$10$^{-3}$)\\

C$_2$H$_3$  &    1.1$\times$10$^{-9}$ (10$^6$)     &  4.7$\times$10$^{-9}$ (10$^6$) & $<$   10$^{-10}$  & 2.1$\times$10$^{-10}$ (7$\times$10$^{-2}$) & $<$   10$^{-10}$  & $<$   10$^{-10}$\\

C$_2$H$_4$  &    8.2$\times$10$^{-8}$ (10$^6$)     &  3.6$\times$10$^{-7}$ (10$^6$ ) &    8.1$\times$10$^{-9}$ (10$^6$)     & 7.1$\times$10$^{-8}$ (9$\times$10$^{-2}$)     & 2.0$\times$10$^{-7}$ (5$\times$10$^{-3}$)     & 7.5$\times$10$^{-7}$ (5$\times$10$^{-3}$)  \\

C$_2$H$_5$  &    8.1$\times$10$^{-10}$ (10$^6$)     &  3.4$\times$10$^{-9}$ (10$^6$) & $<$   10$^{-10}$  & 1.3$\times$10$^{-10}$ (10$^6$)     &  $<$   10$^{-10}$  & 3.6$\times$10$^{-10}$ (2$\times$10$^{-1}$)\\

C$_2$H$_6$  &    5.0$\times$10$^{-9}$ (10$^6$)     &  2.1$\times$10$^{-8}$ (10$^6$) &    4.1$\times$10$^{-9}$ (10$^6$)     &  1.6$\times$10$^{-8}$ (10$^6$)     & 6.2$\times$10$^{-7}$ (2$\times$10$^{-1}$)     & 2.3$\times$10$^{-6}$ (2$\times$10$^{-1}$) \\

aC$_3$H$_4$  &    $<$   10$^{-10}$  & $<$   10$^{-10}$  & $<$   10$^{-10}$  & $<$   10$^{-10}$  & 5.5$\times$10$^{-9}$ (3$\times$10$^{-3}$) & 1.8$\times$10$^{-8}$ (2$\times$10$^{-3}$)  \\

pC$_3$H$_4$  &    $<$   10$^{-10}$  &  1.6$\times$10$^{-10}$ (10$^6$ ) &    $<$   10$^{-10}$  & $<$   10$^{-10}$  & 1.1$\times$10$^{-9}$ (4$\times$10$^{-3}$) & 3.0$\times$10$^{-9}$ (4$\times$10$^{-3}$)\\

C$_3$H$_2$  &    $<$   10$^{-10}$  &  6.1$\times$10$^{-9}$ (2$\times$10$^{-2}$) &    $<$   10$^{-10}$  & 1.7$\times$10$^{-9}$ (6$\times$10$^{-3}$) & 5.3$\times$10$^{-10}$ (2$\times$10$^{-3}$) & 1.1$\times$10$^{-9}$ (2$\times$10$^{-3}$)\\

C$_3$H$_3$  &    $<$   10$^{-10}$  &  2.2$\times$10$^{-9}$ (2$\times$10$^{-2}$) &    $<$   10$^{-10}$  & 7.3$\times$10$^{-9}$ (6$\times$10$^{-3}$) & 2.5$\times$10$^{-9}$ (2$\times$10$^{-3}$) & 5.1$\times$10$^{-9}$ (2$\times$10$^{-3}$) \\

C$_2$H$_3$CHO &    $<$   10$^{-10}$  &    $<$   10$^{-10}$  &    $<$   10$^{-10}$  & $<$   10$^{-10}$   & 3.4$\times$10$^{-10}$ (1$\times$10$^{-3}$) & 8.1$\times$10$^{-10}$ (1$\times$10$^{-3}$)\\

2C$_3$H$_5$ &    $<$   10$^{-10}$  &    $<$   10$^{-10}$  &    $<$   10$^{-10}$  & $<$   10$^{-10}$   & 2.6$\times$10$^{-10}$ (8$\times$10$^{-4}$) & 3.5$\times$10$^{-10}$ (8$\times$10$^{-4}$) \\

C$_3$H$_8$ &    $<$   10$^{-10}$  &    $<$   10$^{-10}$  &    $<$   10$^{-10}$  &  $<$   10$^{-10}$  & 4.6$\times$10$^{-9}$ (1$\times$10$^{-3}$) & 2.9$\times$10$^{-8}$ (2$\times$10$^{-1}$)\\

C$_4$H$_2$  &    $<$   10$^{-10}$  &  1.1$\times$10$^{-10}$ (10$^{2}$) &    $<$   10$^{-10}$  & 1.5$\times$10$^{-9}$ (5$\times$10$^{-3}$) & 3.6$\times$10$^{-9}$ (2$\times$10$^{-3}$) & 6.7$\times$10$^{-9}$ (10$^{-3}$)\\

C$_4$H$_4$O &    $<$   10$^{-10}$  &    $<$   10$^{-10}$  &    $<$   10$^{-10}$  & $<$   10$^{-10}$   & $<$   10$^{-10}$   & 1.4$\times$10$^{-10}$ (1$\times$10$^{-3}$) \\

1C$_4$H$_6$ &    $<$   10$^{-10}$  & $<$   10$^{-10}$  & $<$   10$^{-10}$  & $<$   10$^{-10}$  & 2.0$\times$10$^{-10}$ (3$\times$10$^{-3}$) & 6.9$\times$10$^{-10}$ (2$\times$10$^{-3}$) \\

iC$_4$H$_8$ &    $<$   10$^{-10}$  & $<$   10$^{-10}$  & $<$   10$^{-10}$  & $<$   10$^{-10}$  & $<$ 10$^{-10}$ & 1.4$\times$10$^{-10}$ (3$\times$10$^{-3}$) \\

C$_4$H$_{10}$ &    $<$   10$^{-10}$  &    $<$   10$^{-10}$  &    $<$   10$^{-10}$  &  $<$   10$^{-10}$  & $<$   10$^{-10}$  & 4.4$\times$10$^{-10}$ (2$\times$10$^{-1}$)  \\

cC$_5$H$_6$ &    $<$   10$^{-10}$  &    $<$   10$^{-10}$  &    $<$   10$^{-10}$  & 3.1$\times$10$^{-10}$ (3$\times$10$^{-2}$) & 1.3$\times$10$^{-10}$ (2$\times$10$^{-3}$) & 3.6$\times$10$^{-10}$ (10$^{-3}$)\\

lC$_6$H$_4$ &    $<$   10$^{-10}$  &    $<$   10$^{-10}$  &    $<$   10$^{-10}$  & $<$   10$^{-10}$   & 2.6$\times$10$^{-8}$ (3$\times$10$^{-3}$) & 8.4$\times$10$^{-8}$ (3$\times$10$^{-3}$)\\

cC$_6$H$_6$ &    $<$   10$^{-10}$  &    $<$   10$^{-10}$  &    $<$   10$^{-10}$  & $<$   10$^{-10}$   & 1.1$\times$10$^{-9}$ (2$\times$10$^{-3}$) & 3.1$\times$10$^{-9}$ (2$\times$10$^{-3}$)\\

cC$_6$H$_5$OH &    $<$   10$^{-10}$  &    $<$   10$^{-10}$  &    $<$   10$^{-10}$  & $<$   10$^{-10}$   & 2.2$\times$10$^{-8}$ (3$\times$10$^{-3}$) &  7.6$\times$10$^{-8}$ (3$\times$10$^{-3}$)\\

\hline
\hline 
\end{tabular}
\end{table*}

\subsubsection{Cool ($T_{500}$) profile}
For $T_{500}$, there are differences between the two chemical schemes, both for the solar C/O ratio and for the carbon-rich case. In both cases, important carbon species such as CO and C$_2$H$_2$ are destroyed in the upper atmosphere by photolysis. With the C$_0$-C$_6$ scheme, these species are destroyed to a larger extent than with the C$_0$-C$_2$ scheme, and we found that the "missing" carbon is placed in cC$_6$H$_8$ through the following chemical pathway, starting with the photodissociation of CO: 

\begin{center}
\begin{tabular}{rcl}
CO + h$\nu$ &$\longrightarrow$& C + O($^{3}$P)\\
C + H$_2$ &$\longrightarrow$& CH + H \\
CH + H$_2$ &$\longrightarrow$& $^3$CH$_2$ + H\\
$^3$CH$_2$ + C &$\longrightarrow$& C$_2$H + H\\
C$_2$H + H &$\longrightarrow$& C$_2$H$_2$\\
2 C$_2$H$_2$ &$\longrightarrow$& nC$_4$H$_3$ + H\\
nC$_4$H$_3$ + C$_2$H$_2$ &$\longrightarrow$& lC$_6$H$_5$\\
lC$_6$H$_5$ &$\longrightarrow$& cC$_6$H$_5$\\
cC$_6$H$_5$ + H &$\longrightarrow$& cC$_6$H$_6$\\
cC$_6$H$_6$ + H &$\longrightarrow$& cC$_6$H$_7$\\
cC$_6$H$_7$ + H &$\longrightarrow$& cC$_6$H$_8$\\
\hline
CO + 2 C$_2$H$_2$ +  2 H$_2$ + C &$\longrightarrow$& cC$_6$H$_8$ + O($^{3}$P).\\
\end{tabular}
\end{center}

\noindent Since it involved heavy hydrocarbons, it can only occur for the C$_0$-C$_6$ scheme.

\subsubsection{Warm ($T_{1000}$) and hot ($T_{1500}$) profiles}

As seen in Fig.\ref{fig:avechv}, for any of these two $P$-$T$ profiles there is no difference between the results given by the two chemical schemes when the C/O ratio is solar, but there are some differences in the upper atmosphere when the C/O ratio is $>$ 1. First, we focus on C/O = 1.1.

\begin{figure}
\centering
\includegraphics[angle=0,width=\columnwidth]{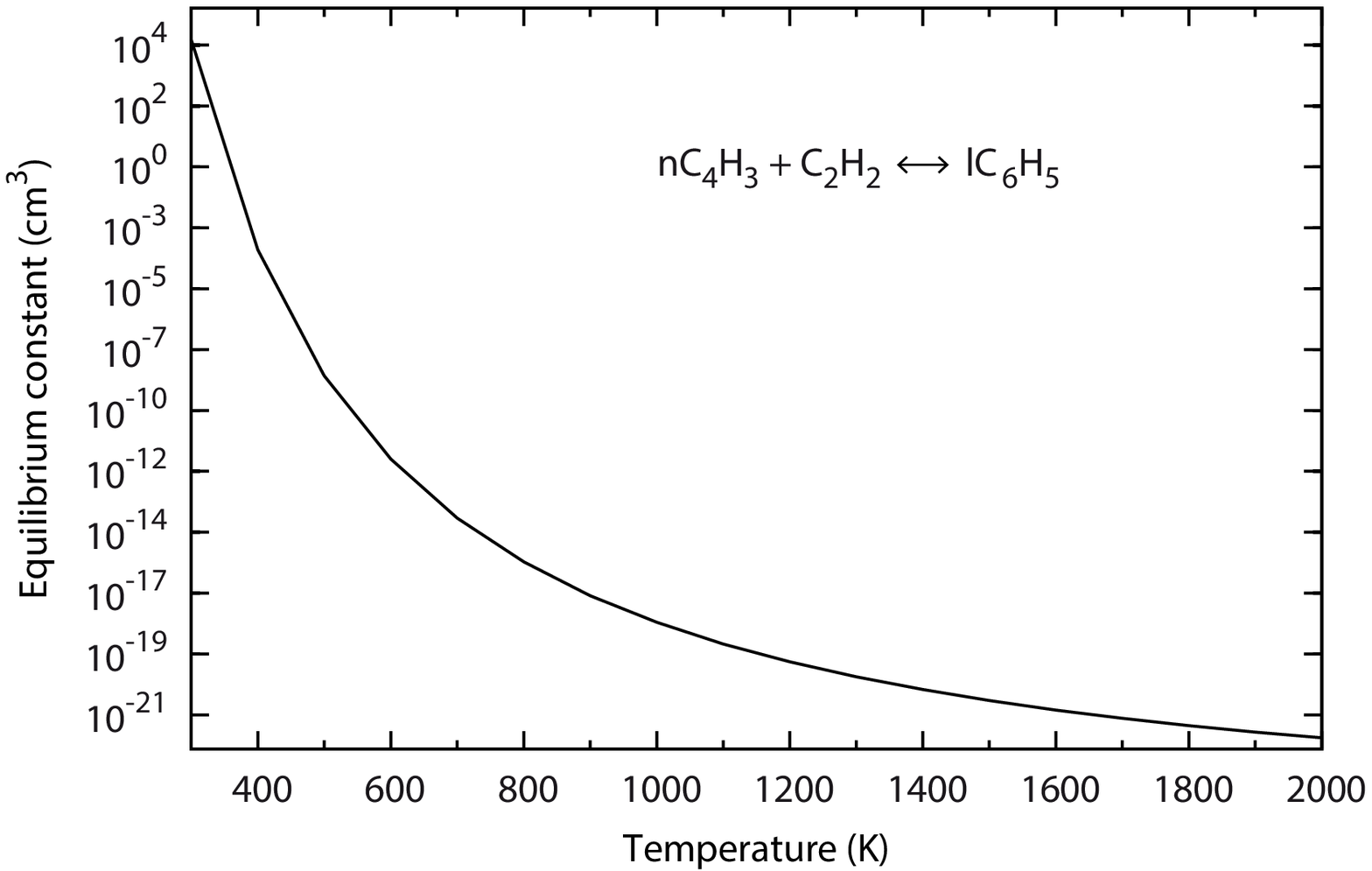}
\caption{Equilibrium constant of the reaction nC$_4$H$_3$~+~C$_2$H$_2$~$\longleftrightarrow$~lC$_6$H$_5$ as a function of temperature. Because the number of reactants and products is different, the equilibrium constant is not dimensionless.}\label{fig:vitesse_lC6H5}
\end{figure}

In the warm and hot profiles, we found that in contrast to the cool profile, cC$_6$H$_8$ is unimportant. We discovered that this heavy species is not abundant when the temperature is high because of the reaction nC$_4$H$_3$ + C$_2$H$_2$ $\longrightarrow$ lC$_6$H$_5$, which is the limiting reaction in the mechanism identified above for the production of cC$_6$H$_8$. The equilibrium constant of this reaction (represented in Fig.~\ref{fig:vitesse_lC6H5}) was calculated as a function of temperature. At low temperature, the equilibrium constant is high so $k_f(T) >>  k_r(T)$, thus, the addition of nC$_4$H$_3$ with C$_2$H$_2$ is favoured to form lC$_6$H$_5$ (and then cC$_6$H$_8$). At high temperature, the equilibrium constant is very low, so the reverse reaction is predominant. The destruction of lC$_6$H$_5$ is favoured. As a consequence, the concentration of cC$_6$H$_8$ is lower.

The main species that show differences in their abundances when using either the C$_0$-C$_2$ or C$_0$-C$_6$ schemes are CH$_3$, CH$_4$, H$_2$O, and CO$_2$, which are more abundant when using the C$_0$-C$_6$ scheme than the C$_0$-C$_2$ scheme. C$_2$H$_2$ and H are also slightly affected and are less abundant for the C$_0$-C$_6$ scheme. This occurs between 10 and 10$^{-3}$ mbar for $T_{1500}$, and between 0.3 and 10$^{-4}$ mbar for $T_{1000}$. We present for each of these species the mechanisms we identified to explain these departures between the two chemical networks.\\

\paragraph{Methyl radical (CH$_3$):}
For CH$_3$, we found that acetylene (C$_2$H$_2$) is responsible for the higher mixing ratio obtained with the C$_0$-C$_6$ scheme. C$_2$H$_2$ forms CH$_3$ through the following chemical pathway, involving heavy hydrocarbons that are only present in the C$_0$-C$_6$ scheme:

\begin{center}
\begin{tabular}{rcl}
C$_2$H$_2$ + H &	$\longrightarrow$ & C$_2$H$_3$\\
C$_2$H$_3$ + H$_2$   &	$\longrightarrow$ & C$_2$H$_4$ + H\\
C$_2$H$_4$ + H &	$\longrightarrow$ & C$_2$H$_5$\\
C$_2$H$_5$ + H &$\longrightarrow$ & 2 CH$_3$\\
				   \hline
C$_2$H$_2$ + 2H + H$_2$ &	$\longrightarrow$& 2 CH$_3$.
				   
\end{tabular}
\end{center}

\noindent This explains the lower abundance of C$_2$H$_2$ when the C$_0$-C$_6$ scheme is used.
 
\paragraph{Methane (CH$_4$):}
Methane is mainly formed (at 99\%) via the reaction CH$_3$ + H$_2$ $\longrightarrow$ CH$_4$ + H, so its abundance is linked to that of CH$_3$. Because CH$_3$ is more abundant for the C$_0$-C$_6$ scheme, there is more methane when using this network.\\


\paragraph{Water (H$_2$O) and carbon dioxide (CO$_2$):}
For the $T_{1000}$ and $T_{1500}$ profiles, H$_2$O and CO$_2$ behave in the same way. These two species are in equilibrium according to the reaction scheme previously found by \cite{moses2011disequilibrium} in hot Jupiters:

\begin{center}
\begin{tabular}{rcl}
H$_2$O + H   & $\longleftrightarrow$ & OH + H$_2$ \\
CO + OH &	$\longleftrightarrow$ & CO$_2$ + H\\
				   \hline
H$_2$O + CO & $\longleftrightarrow$ & CO$_2$ + H.
				   
\end{tabular}
\end{center}

\noindent We remark that H$_2$O and CO$_2$ have different abundances depending on the chemical scheme. These departures are found at $\sim$ 2$\times$10$^{-3}$ mbar for  $T_{1000}$ and around 10$^{-1}$ mbar for $T_{1500}$. In both cases, the two species are more abundant for the C$_0$-C$_6$ than for the C$_0$-C$_2$ scheme.
Identifying why there is more water when using the C$_0$-C$_6$ scheme is very difficult because H$_2$O is involved in very many reactions. Nevertheless, we found a possible explanation: in both chemical schemes, the main reaction (at $\sim$99\%) producing water is OH + H$_2$ $\longrightarrow$ H$_2$O + H, but the reaction OH + CH$_4$  $\longrightarrow$ CH$_3$ + H$_2$O also participates in the formation of water. The contribution of this reaction is two orders of magnitude stronger for the C$_0$-C$_6$ scheme than for C$_0$-C$_2$. Moreover, methane is $\sim$~1000 times more abundant at this level for the C$_0$-C$_6$ scheme. This can probably explain the higher mixing ratio of H$_2$O obtained for the C$_0$-C$_6$ scheme. As we said at the beginning of Sect.~\ref{sec:Resultats}, this hypothesis needs to be confirmed with a specific algorithm, and this will be the subject of a more detailed study. The higher amount of CO$_2$ is then a natural consequence of the equilibrium between the two species.

\begin{figure}
\centering
\includegraphics[angle=0,width=\columnwidth]{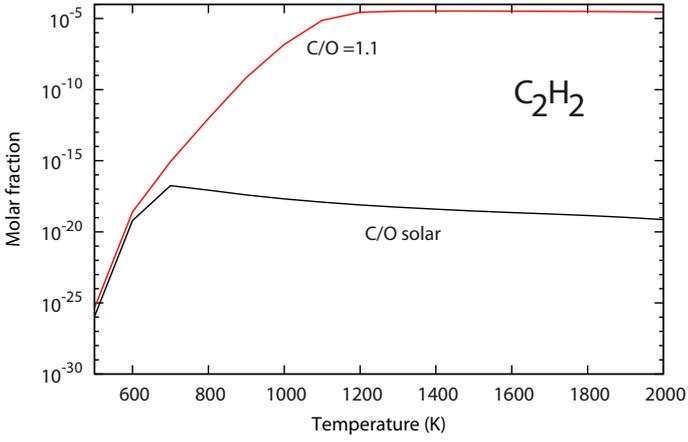}
\caption{Abundance of acetylene at the thermochemical equilibrium as a function of temperature. Two cases are represented: C/O = solar (\textit{black}) and C/O = 1.1 (\textit{red})} \label{fig:C2H2}
\end{figure}

\paragraph{Acetylene (C$_2$H$_2$)}
We have shown that C$_2$H$_2$ is an important species in atmospheres with a high C/O ratio. This species is involved in the formation of CH$_3$ and so is indirectly involved in the formation of CH$_4$, H$_2$O, and CO$_2$. We can see in Fig.~\ref{fig:avechv} that acetylene is a species with a globally  high abundance in the atmosphere for the $\zeta_{1.1}$ cases. But if we compare this with the $\zeta_{0.54}$ cases, we realise that this is different when the C/O ratio is solar. In these cases, C$_2$H$_2$ is not abundant in the atmosphere. Furthermore, C$_2$H$_2$ is also abundant in the $T_{500}$ cases in the upper atmosphere because of photochemistry.\\
As Fig.~\ref{fig:C2H2} shows it, for T $\geqslant$ 800 K, there is a difference of several orders of magnitude between the abundance of C$_2$H$_2$ predicted by thermochemical equilibrium when considering a gas mixture with a solar C/O ratio or equal to 1.1. The low abundance of C$_2$H$_2$ in the $\zeta_{0.54}$ cases explains why there is no departure between the results obtained for the two chemical schemes. If there is no C$_2$H$_2$, the chemical pathways leading to the formation of CH$_3$ identified in the paragraph (CH$_3$) cannot occur. CH$_3$, CH$_4$, H$_2$O, and CO$_2$ will remain at the same abundances as were found for the C$_0$-C$_2$ scheme. This indicates the important role of acetylene in the output of the chemical network calculations.
 
\section{Spectra}\label{sec:Spectra}

To determine whether the differences in abundances obtained using the two chemical schemes have an impact on the spectra, we computed the synthetic spectra corresponding to the 12 atmospheric compositions using the code described in \cite{agundez2014pseudo2D}. For this study, the opacities of C$_2$H$_2$, C$_2$H$_4$, and C$_2$H$_6$ compiled in HITRAN2012 \citep{rothman2013hitran2012} were added because these C-species can potentially be important in C-rich atmospheres under certain conditions. Nevertheless, in the different models we have tested, C$_2$H$_4$ and C$_2$H$_6$ are not sufficiently abundant (always lower than 10$^{-6}$) to affect the synthetic spectra. Note that the opacities of cC$_6$H$_8$ and CH$_3$ are not included because these data are not known, but they could have an important influence on the synthetic spectra and should be added as soon as they are available. For the same reason, opacities of other C$_{n > 2}$H$_x$ species are not included neither. Nevertheless, in view of their low abundances (see Table~\ref{table:abundances_grid}), they would probably not influence the synthetic spectra. We present in this section the spectra in transmission that would be obtained during the primary transit, and then the spectra corresponding to the secondary transit, in emission.

\subsection{Transmission spectra}\label{sec:trans_spectra}

\begin{figure}
\centering
\includegraphics[angle=0,width=0.9\columnwidth]{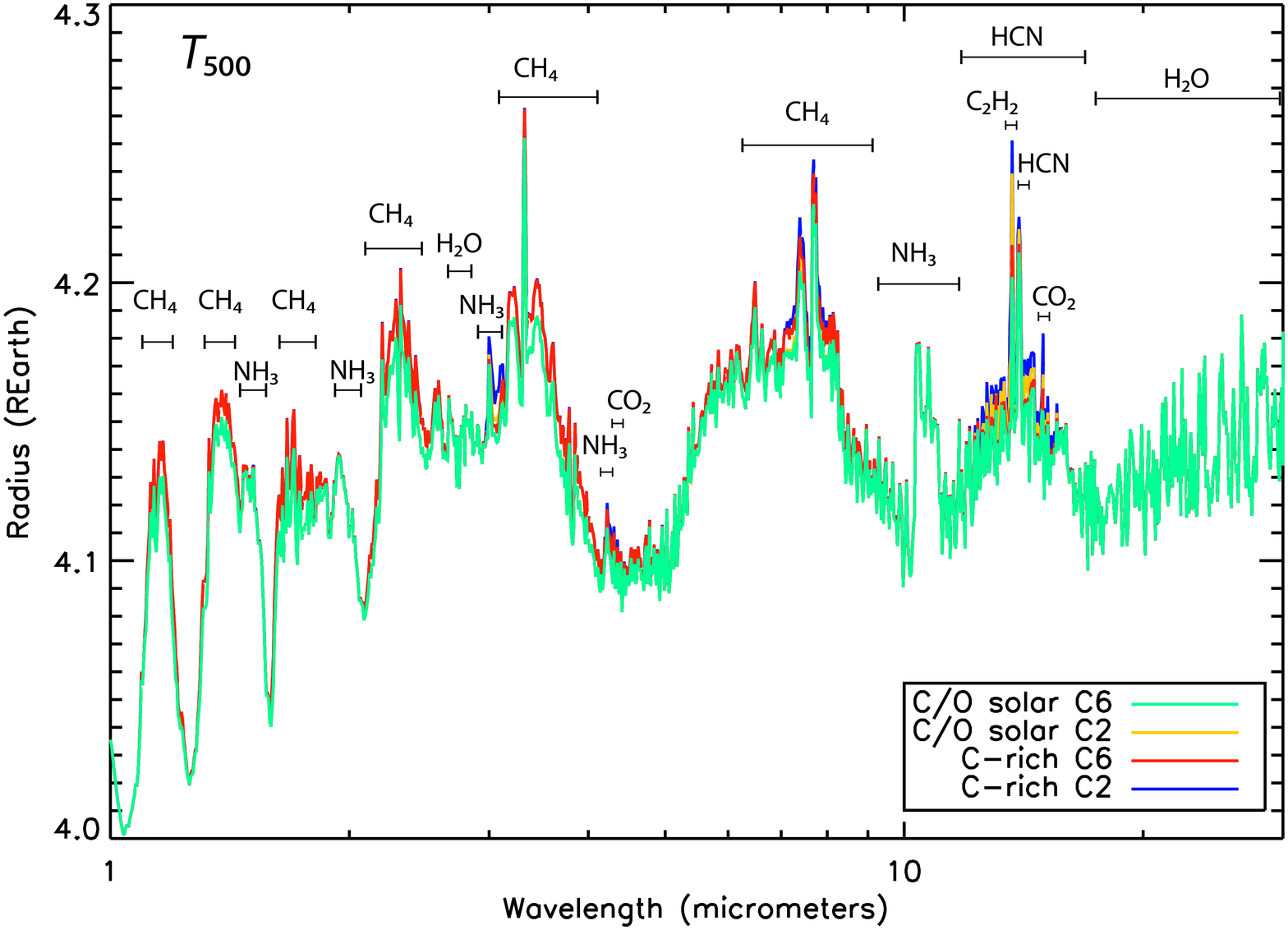}
\includegraphics[angle=0,width=0.9\columnwidth]{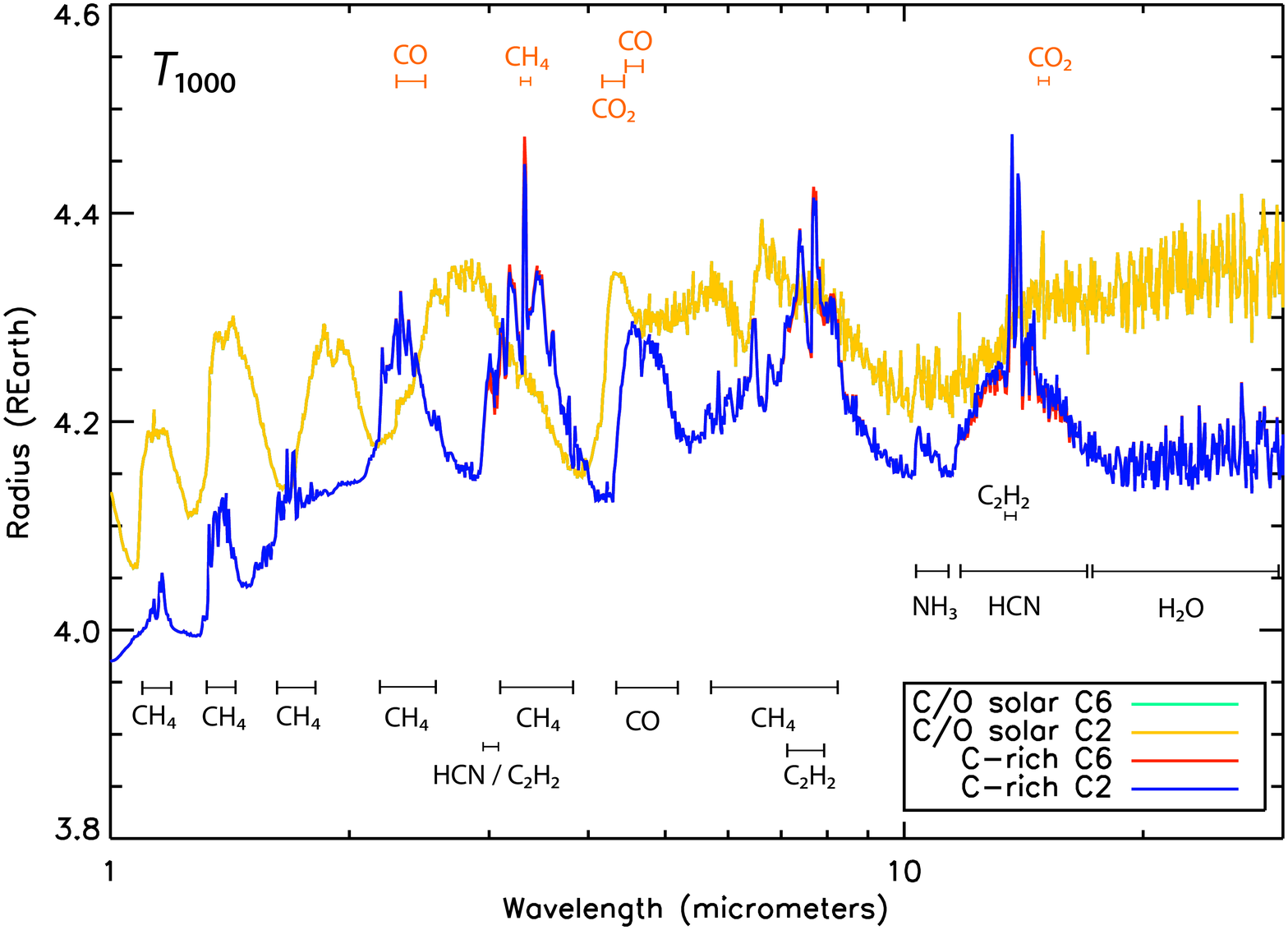}
\includegraphics[angle=0,width=0.9\columnwidth]{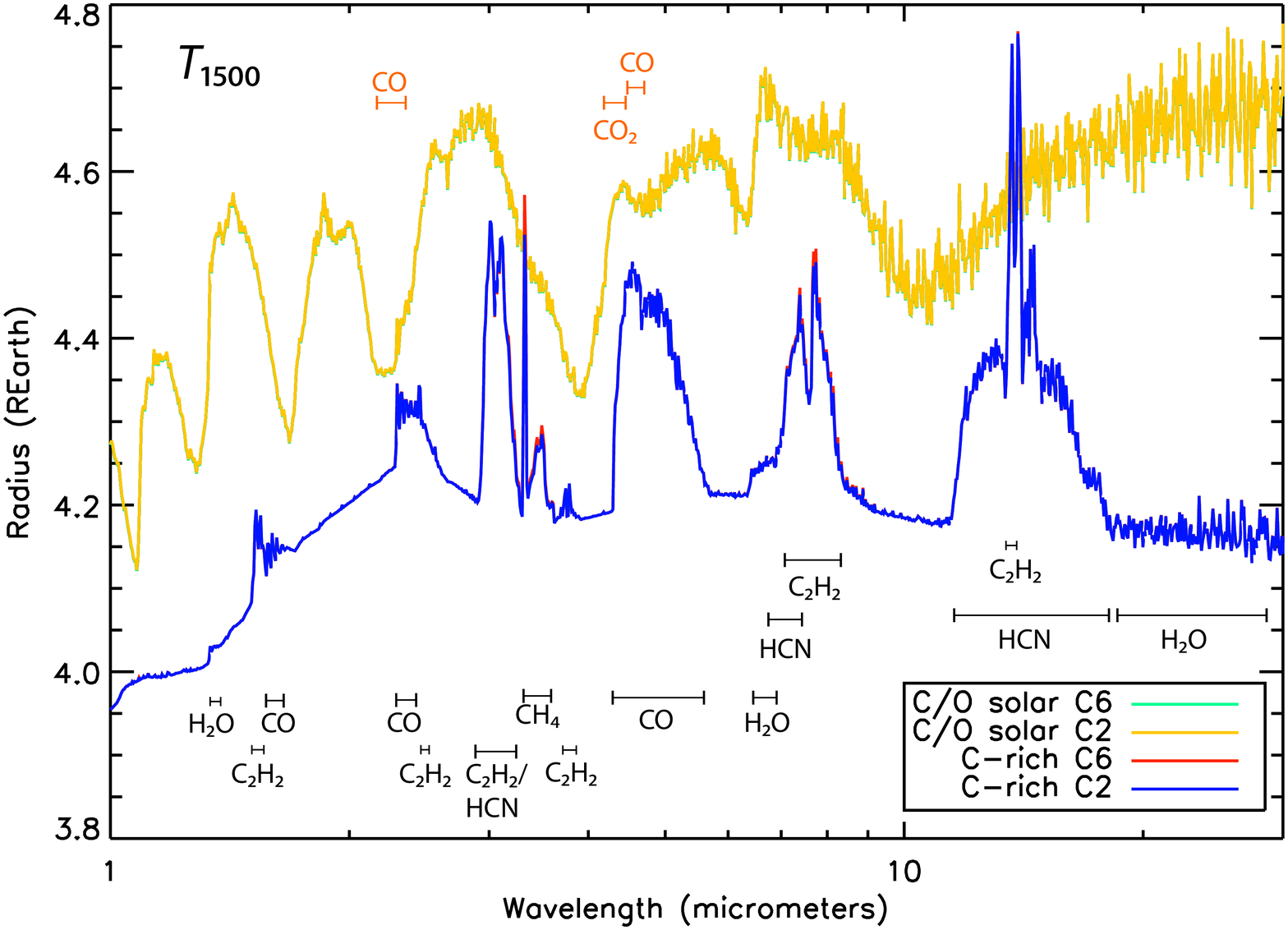}
\caption{Synthetic transmission spectra corresponding to the three thermal profiles $T_{500}$ (up), $T_{1000}$ (middle), and $T_{1500}$ (bottom), represented by the apparent radius. For each thermal profile, spectra have been calculated for the atmospheric compositions found using the C$_0$-C$_2$ and C$_0$-C$_6$ schemes for the cases C/O~=~solar and C/O~=~1.1, as labelled in each figure. Main molecular features are indicated on each spectra. For the $T_{1000}$ and $T_{1500}$ cases, features for the C/O solar spectra are plotted in orange and features for the C-rich spectra in black.} \label{fig:spectra_tra}
\end{figure}

\begin{figure*}[!htb]
\centering
\includegraphics[angle=0,width=0.9\columnwidth]{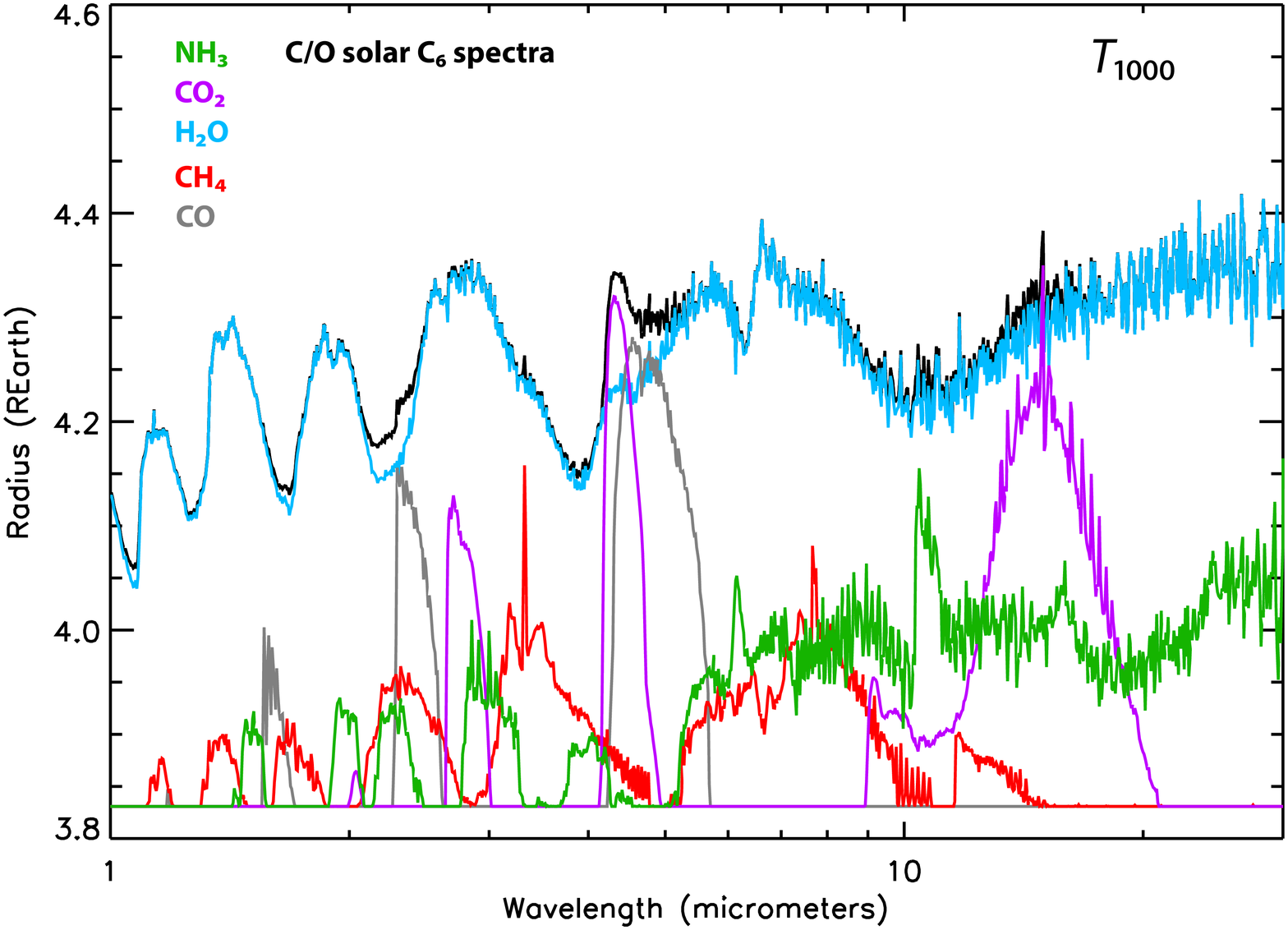}
\includegraphics[angle=0,width=0.9\columnwidth]{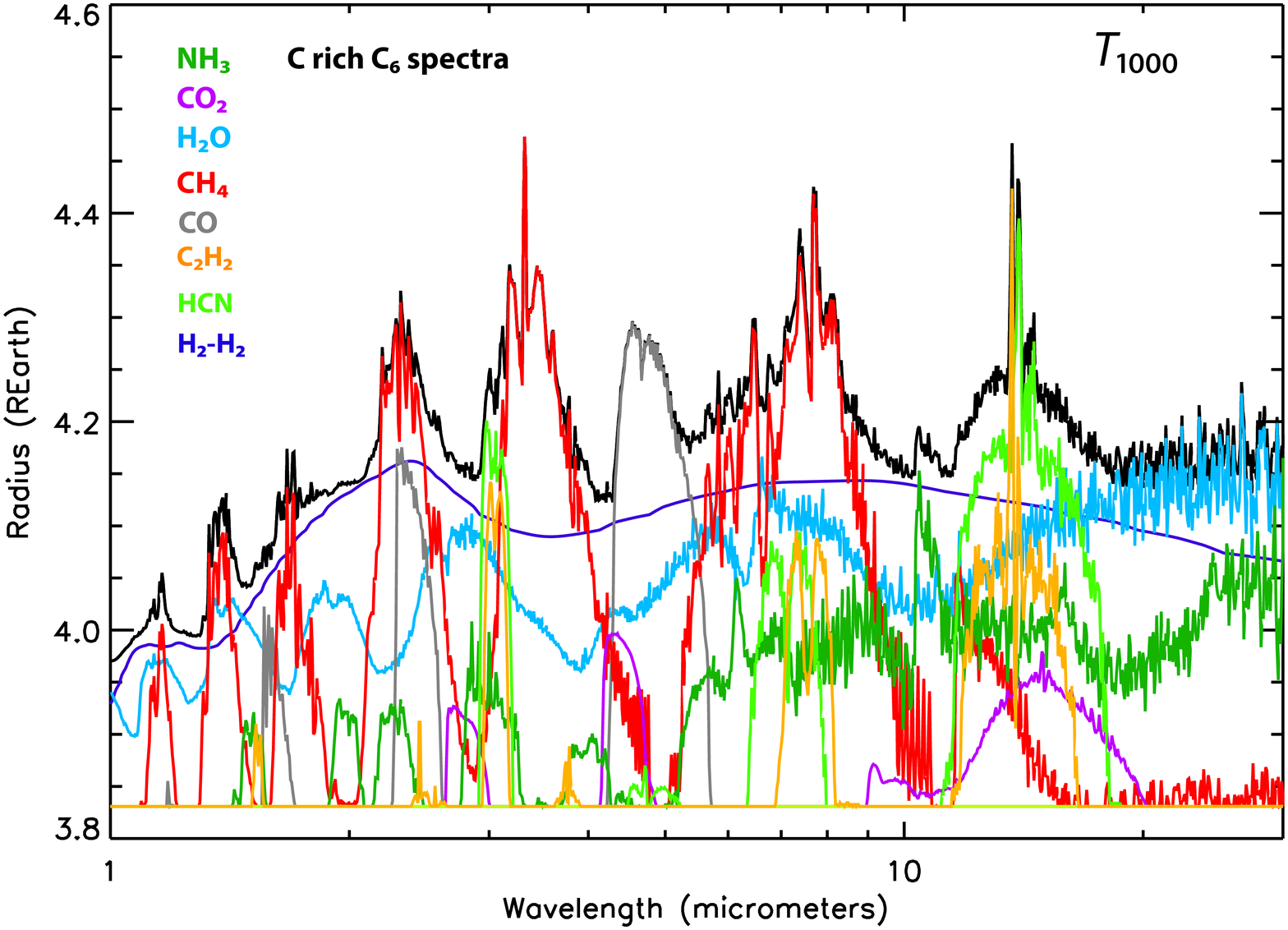}
\caption{Contributions of the main opacity sources to the synthetic transmission spectra corresponding to the thermal profile $T_{1000}$ for the cases C/O~=~solar (left) and C/O~=~1.1 (right). Spectra have been calculated for the atmospheric compositions found using the C$_0$-C$_6$ schemes.} \label{fig:spectra_tra_details}
\end{figure*}

The transmission synthetic spectra computed for the different atmospheres are represented in Fig.~\ref{fig:spectra_tra}. As an example, Fig.~\ref{fig:spectra_tra_details} shows the relative contributions of the main opacity sources for the spectra corresponding to the $T_{1000}$ thermal profile.
 
For the $T_{500}$ profile, all the spectra are very similar because the atmospheric composition is also quite similar for the two C/O ratio cases. The global form of the spectra is due to H$_2$O and CH$_4$. Most of the features are caused by the absorption of these species. In addition, we note the contribution of NH$_3$ around 1.5 and 10.5 $\mu$m. There are two strong peaks beyond 10 $\mu$m. The first one, at 13.6 $\mu$m is due to C$_2$H$_2$, and the second one, very close at 13.9 $\mu$m, is due to HCN. A smaller peak is visible at 14.9 $\mu$m, which is caused by CO$_2$. 
Despite their similarity, there are small departures between the spectra. The differences between the C-rich (green, calculated with the C$_0$-C$_6$ scheme) and C/O solar (red,  calculated with the C$_0$-C$_6$ scheme) spectra are due to methane, which is more abundant in the $\zeta_{1.1}$ case. But there are also differences between spectra corresponding to the two chemical schemes at 3 and 4 $\mu$m, between 7 and 8 $\mu$m, and between 13 and 15 $\mu$m. The departure at 4 $\mu$m is due to the difference of the mixing ratio of CO$_2$ obtained when using the two schemes (blue vs. red for the $\zeta_{1.1}$  case and green vs. yellow for the $\zeta_{0.54}$ case), and all the others are due to C$_2$H$_2$, which is more abundant when using the C$_0$-C$_2$ scheme. This means that this species contributes more to the transmission spectra.

For the $T_{1000}$ profile, we observe strong differences between the spectra corresponding to the C-rich and the C/O solar cases. First, the global form of the C/O solar spectra is due to H$_2$O, with contributions of CO$_2$ around 4.2 and 15 $\mu$m, CO around 2.3 and 4.6 $\mu$m, and CH$_4$ around 2.2 $\mu$m. There is no difference between the spectra corresponding to the results obtained with the C$_0$-C$_2$ and the C$_0$-C$_6$ schemes. 
The two C-rich spectra are quite different from the C/O solar spectra. Their form is mainly due to the H$_2$-H$_2$ collision-induced absorption, with contributions of several species: CH$_4$ around 1.16, 1.38, 1.7, 2.3, 3.3 $\mu$m and between 6 and 9 $\mu$m; HCN around 3 $\mu$m and at 13.9 and 14.2 $\mu$m; CO between 4.3 and 5.2 $\mu$m; C$_2$H$_2$ around 3 $\mu$m and at 7.4, 7.7, and 13.6 $\mu$m; NH$_3$ around 10.5 $\mu$m. HCN is the absorber that generates the features from 11 to 16 $\mu$m, and H$_2$O is the main absorber from 16 to 30 $\mu$m.
Unlike the $\zeta_{0.54}$ case, the two $\zeta_{1.1}$ spectra corresponding to the two chemical schemes present small departures. We found these deviations between 3 and 4 $\mu$m, between 7 and 8 $\mu$m, and between 13 and 15 $\mu$m. Like for the $T_{500}$ spectra, these departures are caused by the change of the abundance of C$_2$H$_2$ between the two chemical schemes, but also due to the lower mixing ratio of methane that we obtain when using the C$_0$-C$_6$ scheme.

For the $T_{1500}$ profile, we also observe departures between the spectra corresponding to the two C/O ratios. For the C/O solar spectra, the contributing species are the same as for the $T_{1000}$ profile: H$_2$O everywhere except around 2.3 and 4.6 $\mu$m (CO) and around 4.2 $\mu$m (CO$_2$). The peak of CO$_2$ at 15 $\mu$m is not visible because of the very much lower abundance of this species in the atmosphere. This area of the spectra is entirely dominated by water. We note that once again, there is no difference between the spectra corresponding to the two chemical schemes, which is expected because there is also no difference in the computed chemical abundances.
The spectra of the C-rich atmosphere are different from the C/O solar spectra. Like for the $T_{1000}$ profile, their form is due to the H$_2$-H$_2$ collision. There are contributions of H$_2$O around 1.35 $\mu$m and from 18 $\mu$m until the end of the computed spectra (30 $\mu$m). There are many contribution of C$_2$H$_2$: around 1.5, 2.45, 3, 3.7, 7.4, 7.7, and the very high peak at 13.6 $\mu$m. We see CO features at 1.6, 2.3, and 4.6 $\mu$m. The contribution of CH$_4$ is very small compared to the two previous cases ($T_{500}$ and $T_{1000}$). There is a strong peak at 3.32 $\mu$m and a smaller peak around 3.49 $\mu$m. Methane is also responsible for the absorption peak at 7.39 $\mu$m, which is almost mingled with the two strong acetylene peaks. Finally, the last contributing molecule is HCN with its peaks around 3, 7, 13.9, and 14.2 $\mu$m. HCN is also responsible for the features from 11 to 18 $\mu$m. These results agree with what has been found for WASP-12b by \cite{kopparapu2012photochemical}.
 
\subsection{Emission spectra}

\begin{figure}
\centering
\includegraphics[angle=0,width=0.9\columnwidth]{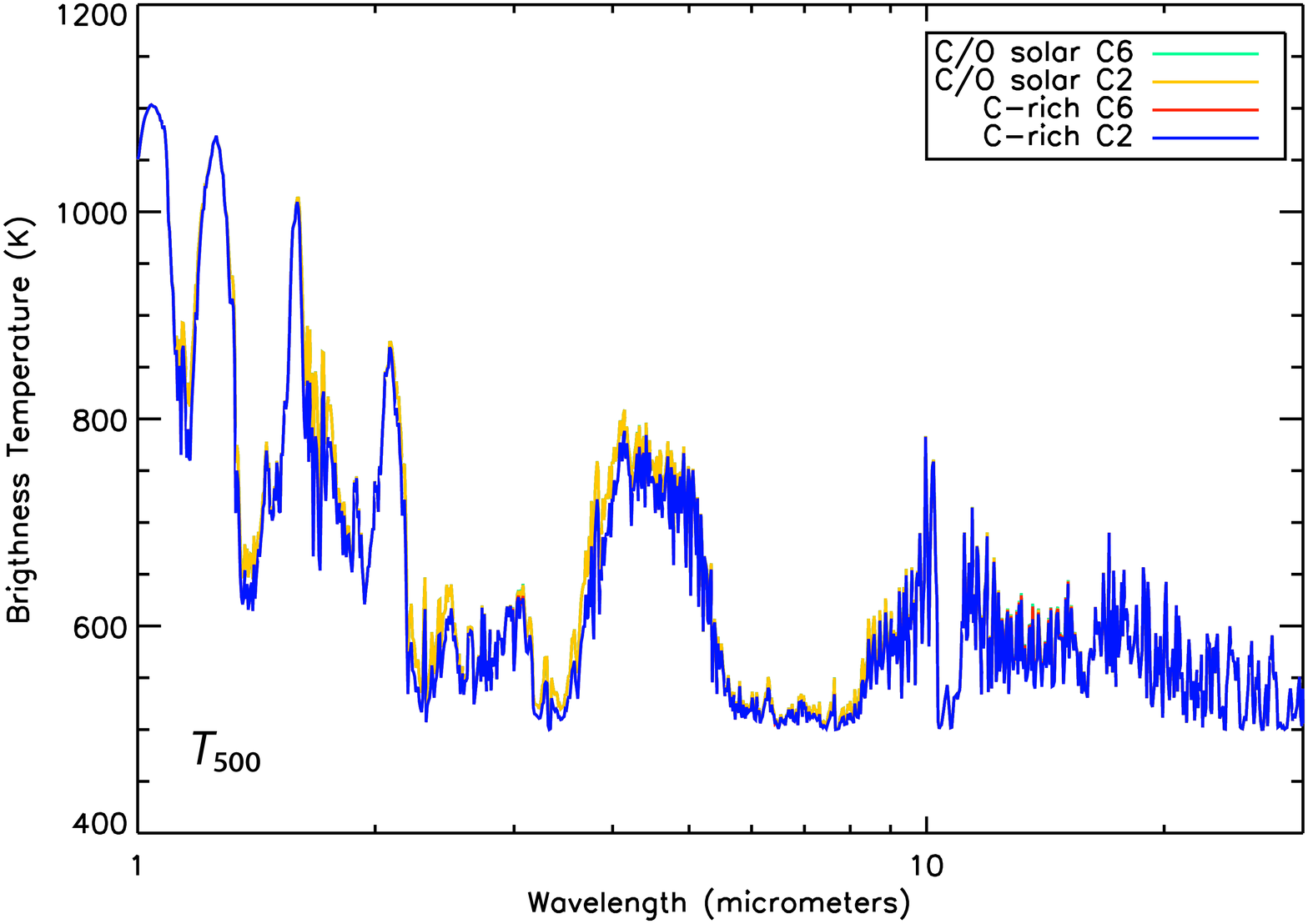}
\includegraphics[angle=0,width=0.9\columnwidth]{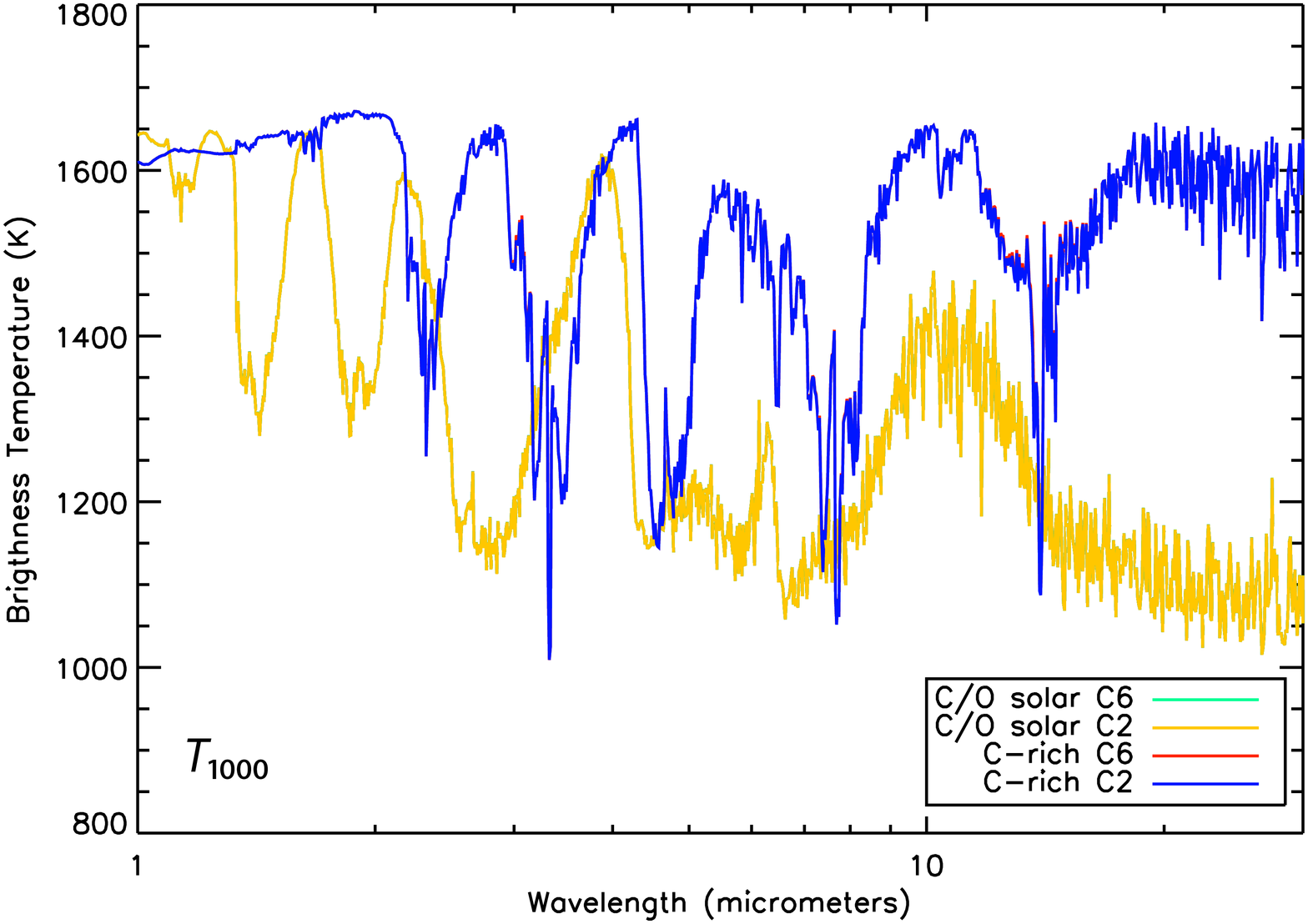}
\includegraphics[angle=0,width=0.9\columnwidth]{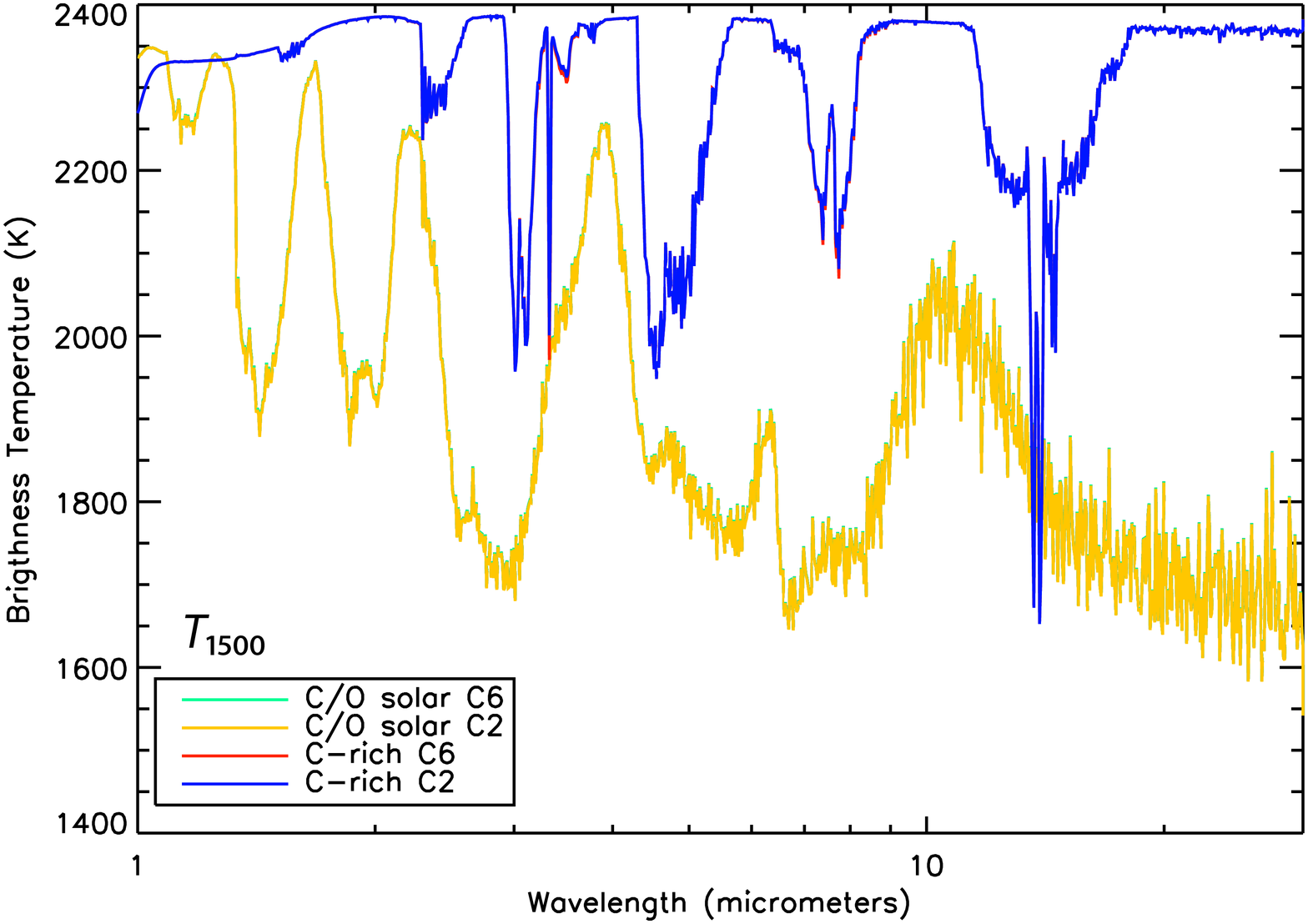}
\caption{Synthetic emission spectra corresponding to the three thermal profiles $T_{500}$ (up), $T_{1000}$ (middle), and $T_{1500}$ (bottom), represented by the brightness temperature (K). For each thermal profile, spectra have been calculated with the atmospheric compositions found with the C$_0$-C$_2$ and C$_0$-C$_6$ schemes for the cases C/O = solar and C/O = 1.1, as labelled in each figure.} \label{fig:spectra_em}
\end{figure}

Figure \ref{fig:spectra_em} presents the synthetic emission spectra obtained for our 12 atmospheres. As for the transmission spectra, for the cool atmosphere, there is almost no difference between the emission spectra of the two different C/O ratios, whereas for the two warmer profiles the emission spectra are clearly different depending on the C/O ratio of the atmosphere.
For the C/O solar cases, the spectra are dominated by the absorption of H$_2$O for the two warmer cases and by water and methane for the $T_{500}$ profile. For the C-rich cases, as we explained for transmission spectra in Sect.~\ref{sec:trans_spectra}, the spectra are dominated by the H$_2$-H$_2$ collision, with contributions of other species, such as C$_2$H$_2$, HCN, and CH$_4$.

The analysis of these emission and transmission spectra shows us that C$_2$H$_2$ and HCN can be considered as good tracers of the C/O ratio if the atmosphere presents a sufficiently high temperature (above 800~K), thanks to their peaks at 7.4, 7.7, 13.6, 13.9, and 14.2 $\mu$m, respectively, visible when C/O = 1.1. For lower temperatures, acetylene and hydrogen cyanide present similar mixing ratios in the two C/O cases that we tested and thus cannot give information on the C/O ratio of the atmosphere. As emphasised by \cite{Madhu2012}, these two species absolutely need to be considered in spectra models to interpret and understand observations of exoplanets atmospheres.

\section{Conclusions}

We have developed a new chemical scheme for studying C-rich atmospheres. This C$_0$-C$_6$ scheme describes the kinetics of 240 species that contain up to six carbon atoms and are involved in 4002 reactions. It is available in the online database KIDA: Kinetic Database for Astrochemistry \citep{KIDA2012} at http://kida.obs.u-bordeaux1.fr/models/. We used this chemical scheme to study atmospheres with different thermal profiles and C/O ratios and compared the results obtained with those obtained with a smaller chemical scheme (C$_0$-C$_2$, described in \citealt{venot2012chemical}). When photodissociation is neglected, the two chemical schemes yield identical results. This strengthens the robustness of the small C$_0$-C$_2$ scheme. It perfectly described the kinetics of C$_2$ species, and even a scheme that contains twice more reactions gives the same results.\\
The introduction of the photodissociations induces, in some cases, differences between the two chemical schemes. With the warm and hot profiles ($\sim$1000~K and $\sim$1500~K), there are differences when the C/O ratio of the atmosphere is high (1.1), but not when it is solar. This is mainly due to C$_2$H$_2$, which is much more abundant in a C-rich atmosphere and participates in the C$_0$-C$_6$ scheme in formation pathways involving heavy hydrocarbons.
For the cool profile ($\sim$500~K), there are differences between the two schemes regardless of the C/O ratio of the atmosphere. Here again, the high amount of C$_2$H$_2$ in the upper atmosphere is responsible for this.
Nevertheless, the differences obtained using the two chemical schemes do not create differences in the synthetic spectra. The analysis of the synthetic spectra corresponding to the 12 different cases has highlighted the fact the absorption features around 7 and 14 $\mu$m due to C$_2$H$_2$ and HCN can be used as tracers of the C/O ratio for atmospheres with temperatures higher than 800 K.\\

In conclusion, our advice for using of one or the other chemical scheme is as follows: \\
- In warm atmospheres (T $\geq$1000K) with solar C/O ratios, the C$_0$-C$_2$ scheme is sufficient.\\
- In cooler atmospheres (regardless of the C/O ratio) or warm atmospheres with a C/O ratio higher than 1, the choice of the chemical scheme depends on the goal of the study. If the focus is on computing synthetic spectra, then the use of the C$_0$-C$_2$ scheme is a reasonable choice because the computation time will be shorter. But if the chemical composition is to be studied in detail and the chemical pathways occurring in the atmosphere are to be understood, then using the more complete C$_0$-C$_6$ scheme is advised.

We recall that the mixing ratios calculated here could change by several orders of magnitude if absorption cross-sections at high temperatures are used for all the absorbing species present in the chemical network. We only used hot data for CO$_2$ \citep{venot2013high} and NH$_3$ \citep{venot_NH3}, but measurements of other species are much needed. In particular, this study shows that C$_2$H$_2$ should be the next target of these experiments.

\begin{acknowledgements}

O.V. acknowledges support from the KU Leuven IDO project IDO/10/2013 and from the FWO Postdoctoral Fellowship programme. The authors thank I. Baraffe, M. Dobrijevic, and F. Selsis for useful discussions. We also thank the anonymous referee and for comments that much improved the manuscript.

\end{acknowledgements}

\bibliographystyle{aa}
\bibliography{bib_C0C6}

\end{document}